# Proton-electron coupled catalyst for ionomer-free electrochemical energy conversion

**Authors:**

Ao Zhang[1], Ran Wang[2], Mohammed O. Bazaid[3], Shiyi Wang[4], Ting-Jung Hsiao[1], Yibo Wang[1], Antonio Sorrentino[5], Yang Liu[1], Yu-Han Joseph Tsai[1], Boxuan Zhou[1], Bosi Peng[2], Zeyan Liu[1], Xiangfeng Duan[2, 6], Adam Z. Weber[4], William A. Goddard III[7], Seung Soon Jang[3], and Yu Huang[1, 2, 6]

**Affiliations:**

[1] Department of Materials Science and Engineering, University of California, Los Angeles, CA 90095, USA

[2] Department of Chemistry and Biochemistry, University of California, Los Angeles, CA 90095, USA

[3] School of Materials Science and Engineering, Georgia Institute of Technology, GA 30332–0245, USA

[4] Energy Technologies Area, Lawrence Berkeley National Laboratory, Berkeley, CA 94720, USA

[5] Max Planck Institute for Dynamics of Complex Technical Systems, 39106 Magdeburg, Germany

[6] California NanoSystems Institute, University of California, Los Angeles, CA 90095, USA

[7] Materials and Process Simulation Center, California Institute of Technology, CA 91125, USA

Corresponding Author. Email: yhuang@seas.ucla.edu, seungsoon.jang@mse.gatech.edu, wag@caltech.edu

**Abstract:**

Efficient electrochemical energy devices are vital to renewable energy technology, yet coordinating the effective flow of electrons, ions, and chemical species continues to be a major challenge. In conventional proton-exchange membrane fuel cell (PEMFC) catalyst layers, proton and electron transport are supplied separately through percolating carbon networks and ionomer binders, rendering the catalyst largely passive and imposing fundamental trade-offs between reactant accessibility, ionic conductivity, and catalyst activity. Here, we introduce a one-dimensional proton-electron coupled catalyst (PECC) design, a transport-integrated electrocatalyst architecture in which the catalyst itself simultaneously supplies electronic and protonic transport to catalyst active sites. Using this PECC, PEMFCs can have an ionomer-free cathode catalyst layer (CCL), resulting in a dramatic 95% reduction in non-Fickian oxygen transport and boosting power density by 34% and 85% compared to traditional CCLs, with cathode Pt loadings of approximately 0.090 mg/cm$^2$ and 0.037 mg/cm$^2$, respectively. Meanwhile, PECC retains 65% of its mass activity and exhibits 32% higher power density than its ionomer-based CCL counterpart after 30k accelerated stressed test. Similar mass transport improvements have been observed in the electrochemical hydrogen pump (EHP) using PECC in the catalyst layers. Molecular dynamics simulations show the PECC's proton conductivity is 249% higher than Nafion. This PECC catalyst structure addresses core transport problems in PEMFCs, leading to almost 20% improvement in fuel efficiency and opens up new possibilities for designing high-performance, cost-effective electrochemical devices.

## Main Text:

Electrochemical reactions that convert electrical and chemical energies are key elements in the renewable energy economy.[1] The efficiency of electrochemical reactions depends not only on the activity of the electrocatalysts used but also critically on the efficient shuttling of electrons, ions, and reactant molecules to the active catalytic sites. The ideal design of an electrocatalyst system should allow simultaneous free access of the catalyst surface to ions, electrons, and chemical molecules, which unfortunately remains challenging.[2-4] In proton-exchange membrane fuel cells (PEMFCs), for example, the ionomer (the polymer electrolyte, e.g., Nafion) used for transporting protons to the platinum (Pt) catalyst simultaneously blankets its surface, hindering the oxygen access at the cathode where the sluggish oxygen reduction reaction (ORR) occurs.[5,6] These limitations can reduce reaction efficiency and limit the overall system output. Similar mass transportation challenges have been reported for various electrochemical systems due to the restricted access of reactants, such as $H_2$ and $CO_2$, to the electrocatalysts.[4,7,8]

Efficiency and performance constraints arising from mass transport are further exacerbated by lower catalyst loading, which is necessary to minimize system costs when employing precious metals such as Pt.[9,10] In case of PEMFCs, the U.S. Department of Energy (DOE) targets a practical Pt total loading $\leqslant$ 0.125 $mg_{Pt\text{-}total}/cm^2$ with high power performance $\geqslant$8 $kW/g_{Pt\text{-}total}$,[11] while ideally a cathode Pt loading $\leqslant$ 0.05 $mg_{Pt\text{-}cathode}/cm^2$ that delivers a performance $\geqslant$20 $kW/g_{Pt\text{-}cathode}$ has been proposed[9,10] but remain unattainable.

Achieving demanding system performance and cost targets requires simultaneous improvement in catalyst activity and mass transport. In PEMFC, over the past few decades, considerable efforts have led to nearly a tenfold increase in the mass activity (MA) of Pt-based ORR catalysts.[12-23] However, the corresponding improvements in cell power density have been modest, falling short of expectations relative to the enhanced catalytic activity. In conventional PEMFC catalyst layers (CLs), the ionomer covers over 75% of the Pt surface, forming a polymer layer between 4 - 17 nm[5,6] (Supplementary Table 1) and introducing substantial local gas transport resistance. Particularly, the ionomer-induced oxygen transport resistance in the cathode catalyst layer (CCL) alone accounts for more than 40% of the total mass transport resistance[9] and significantly limits oxygen supply to catalytic sites (Extended Data Fig.1).[24] Numerous strategies have been proposed to mitigate this challenge, particularly for PEMFC cathodes, including modifying the ionomer molecular structure,[25,26] engineering pores of carbon,[27-29] lowering the ionomer content,[30,31] adjusting distribution,[32,33] incorporating oxygen-transport additives,[34-36] and patterning the CL morphology.[37,38] However, in these designs, the catalyst–ionomer interface persists, as does the dominant local oxygen transport resistance. Attempted ionomer-free CLs suffer from orders-of-magnitude lower proton conductivity that inevitably compromises the device performance.[39-41]

In this study, we present a one-dimensional proton–electron coupled catalyst (PECC) design, a transport-integrated electrocatalyst architecture in which the PECC simultaneously supplies electronic and protonic transport to active sites. Using this PECC, PEMFCs can have an ionomer-free cathode catalyst layer (CCL), wherein efficient delivery of electrons, ions, and reactants to the active sites can sustain high reaction rates. (Fig. 1a). The PECC design featuring a one-dimensional (1D) electron-conductive network based on carbon nanotubes (CNTs), a chemically functionalized CNT surface for ionic transport (termed IonoSkin), and an ionomer-free catalyst surface for oxygen access (Fig. 1a,b). The PECC CCL delivers an impressive **~30-fold enhancement in local $O_2$ mass transport efficiency** (Fig. 1c), **a 2.5-fold increase in proton conductivity** (Fig. 1d), **and a 2.2-fold increase in the electron conductivity** (Fig. 1e), which together demonstrates much higher power performance compared to the conventional CCL, delivering a peak power density of 1.51 W/cm$^2$ under $H_2$/Air conditions at a Pt loading of ~0.090 mg$_{Pt\text{-}cathode}$/cm$^2$. Even at an ultralow Pt loading of 0.037 mg$_{Pt\text{-}cathode}$/cm$^2$, the **PECC attains a 1.26 W/cm$^2$ peak power, 85% higher than the conventional counterpart (Fig. 1f), and a rated power of 22.2 kW$_{Rated}$/g$_{Pt\text{-}cathode}$ that exceeds the industry's stretch target of >20 kW$_{Rated}$/g$_{Pt\text{-}cathode}$.** The PECC has also been applied in the anode (hydrogen oxidation reaction (HOR)) and the cathode (hydrogen evolution reaction (HER)) CLs in an electrochemical hydrogen pump (EHP) with ultralow Pt loadings (0.01 mg$_{Pt}$/cm$^2$), where it greatly improved mass transport properties and reduced the required cell voltage to reach 1 A/cm$^2$ (Fig. 1g,h), indicating a generalizable strategy to mitigate mass-transport limitations in emerging electrochemical technologies.

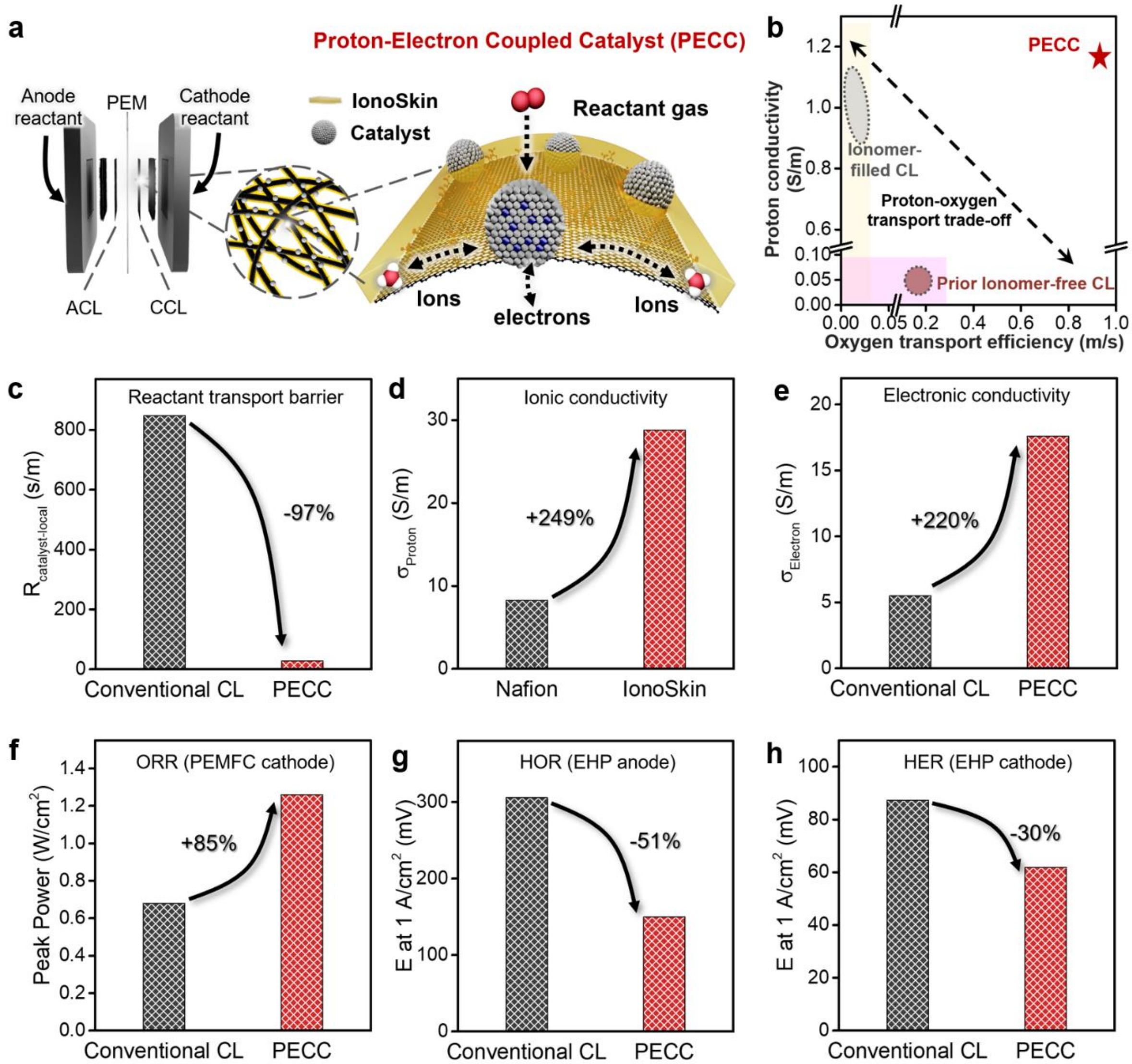

**Fig. 1. | Proton–electron coupled catalyst (PECC) design. a,b**, the **Design Schematic.** The PECC comprises a 1D electron-proton co-conducting architecture design, wherein the electrons are transported to the catalysts through the core of the CNTs and the protons are transported through the IonoSkin (yellow surface layer), leaving the catalyst surface open to the reaction gas. The PECC design eliminates the catalyst-ionomer interface and greatly enhances the performance of an electrochemical cell. **b**, Conventional CLs are oxygen-limited, while prior ionomer-free CLs are proton-limited. This reflects an intrinsic proton–oxygen trade-off, which PECC overcomes by enabling both. **c**, **Reactant Transport:** PECC reduces local oxygen transport resistance ($R_{local}$) by 97% when compared to a conventional CL design in PEMFC; **d**, **Proton Transport**: The PECC demonstrates 249% higher proton conductivity

than that of Nafion under comparable conditions. **e**, **Electron Transport**: The PECC demonstrates 220% higher proton conductivity than that of conventional ionomer percolated carbon CL. **f**, **Power improvement in PEMFC**: PECC leads to an 85% peak power density improvement compared to conventional CL in PEMFC at 0.04 $mg_{Pt\text{-}cathode}/cm^2$. **g**,**h, Cell voltage reduction in electrochemical hydrogen pump (EHP):** At 1 $A/cm^2$ and Pt loading 0.01 $mg_{Pt}/cm^2$, PECC CLs reduced the cell voltage requirement by 51% (HOR, EHP anode, **g**) and 30% (HER, EHP cathode, **h**), respectively.

**Synthesis and characterization of the PtNi/PECC**

The PtNi nanocatalysts (NCs) were used as the model catalyst, which were first grown on the surface of multi-walled CNTs through the impregnation-annealing method (Method 2.1).[42-44] Here, the CNT is selected for its high electron conductivity and inherent 1D morphology as continuous transport pathways. In a typical process, platinum (II) acetylacetonate ($Pt(acac)_2$) and nickel (II) acetylacetonate ($Ni(acac)_2$) were mixed in acetone with stirring. The mixture was then dried and annealed at 400 °C under an argon atmosphere, followed by an acid-wash and annealing treatment to produce the final catalyst, labeled as "**PtNi/CNT**". Then, the sulfonic acid groups ($-SO_3H$) were introduced to the surface of CNTs by the diazonium reaction, transforming the CNT support to the PECC support (Fig. 2a). The final catalytic structure is termed "**PtNi/PECC**". Powder X-ray diffraction (XRD) studies of PtNi/PECC show a face-centered cubic structure diffraction pattern, highly similar to that of PtNi/CNT (Fig. 2b), indicating that the size of PtNi NCs is maintained during the diazonium reaction. The X-ray photoelectron spectroscopy (XPS) (Fig. 2c) of the PtNi/PECC shows distinct S 2p spectrum peaks corresponding to the sulfonic group features, confirming successful grafting of sulfonic groups in PtNi/PECC. The transmission electron microscopy (TEM) image of PtNi/PECC (Fig. 2d) shows well-dispersed PtNi NCs with a small and uniform particle size of 2.48 ± 0.83 nm (Fig. 2d inset). agreeing with the PtNi NC size calculated from XRD (2.41 nm). Inductively coupled plasma atomic emission spectroscopy (ICP-AES) shows that PtNi/PECC contains 20.2 wt% Pt with a Pt: Ni atomic ratio of 81.5:18.5, closely matching PtNi/CNT and c-PtNi/C (Supplementary Table 2). Scanning Transmission Electron Microscopy (STEM) Energy Dispersive X-ray Spectroscopy (EDS) mapping shows the PtNi NC has a Pt-rich shell (Extended Data Fig. 2). In addition, the STEM-EDS revealed a uniform distribution of S across the PECC (Fig. 2e,f) with an S-to-C weight ratio of 1.04 wt.% (Supplementary Table 3).

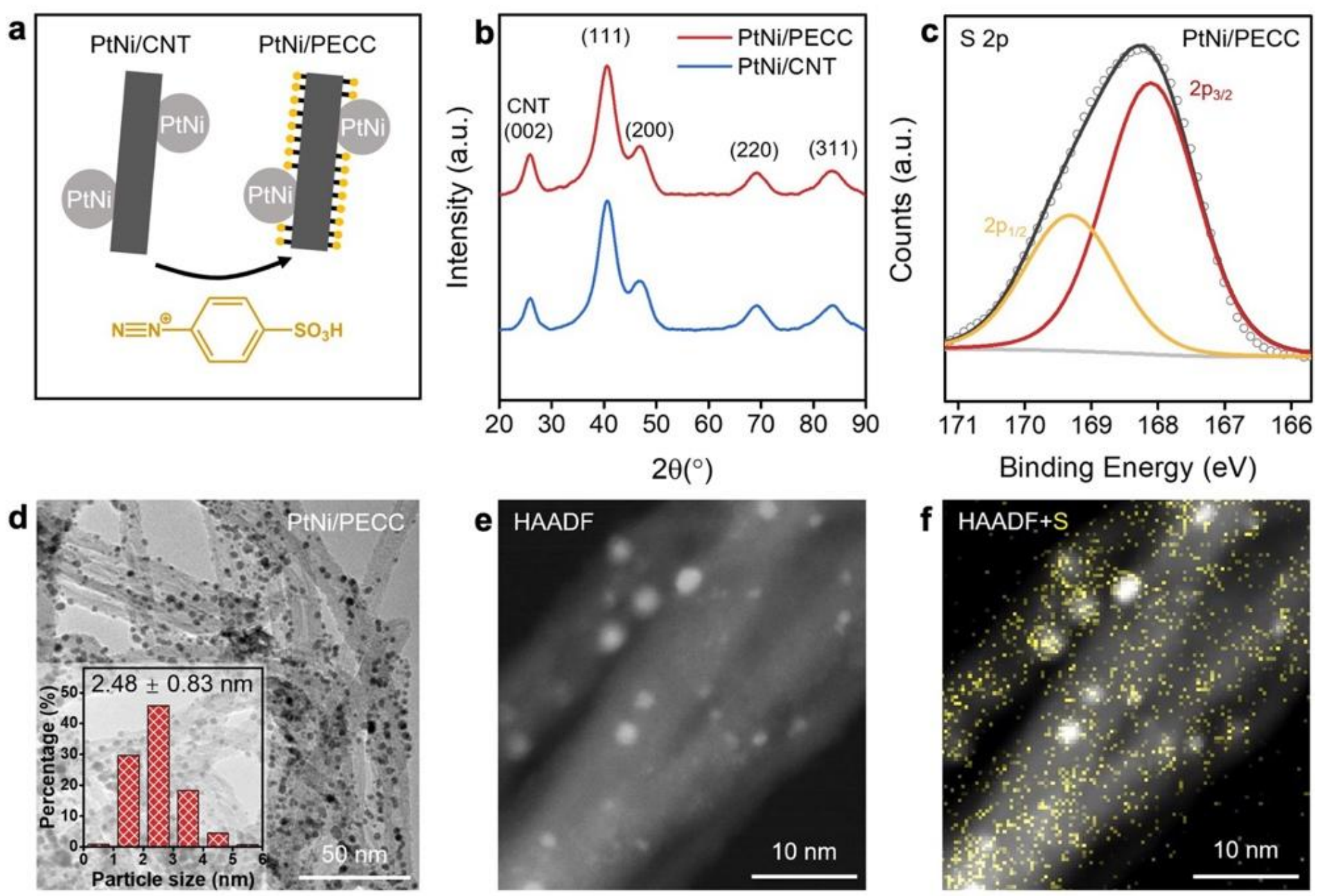

**Fig. 2. | Synthesis and characterization of the PtNi/PECC. a**, Schematic illustration of the PtNi/PECC synthesis. **b**, XRD patterns of PtNi/PECC and PtNi/CNT. **c**, XPS S 2p spectrum confirming the sulfonic functionalization. **d**, TEM image with corresponding PtNi nanoparticle size distribution (inset). **e**, High-angle annular dark-field (HAADF) STEM image. **f**, Superimposed HAADF and the corresponding STEM-EDS mapping of S shows a uniform distribution of S.

## The PEMFC performance

The PEMFC performance of the **PtNi/PECC** CCLs was evaluated in a 5-$cm^2$ membrane electrode assembly (MEA) and compared to MEAs containing commercial PtNi/C and ionomer (termed "**c-PtNi/C+I**"), and PtNi/CNT and ionomer ("**PtNi/CNT+I**") (Supplementary Table 2). To evaluate the performance of the PtNi/PECC under DOE loading target (≤ 0.125 $mg_{Pt\text{-}total}/cm^2$) and under challenging ultralow Pt loading targets, we prepared the MEAs with three distinct Pt loadings of ~0.090, 0.060, and 0.040 $mg_{Pt\text{-}cathode}/cm^2$ at the cathode. Together with the anode loading of 0.020 $mg_{Pt\text{-}anode}/cm^2$, the total Pt loading was 0.110, 0.080, and 0.060 $mg_{Pt\text{-}total}/cm^2$, respectively. **The total Pt loading of 0.060 $mg_{Pt\text{-}total}/cm^2$ represents an ultralow Pt loading that meets the ambitious industrial stretch target (0.0625 $mg_{Pt\text{-}}$**

**$_{total}$/cm$^2$)**, supporting broad PEMFC deployment while easing Pt supply pressure.[9] For a fair comparison, all MEAs used identical components except for the CCL.

The MA of PtNi catalysts in three different CCLs was first examined at cathode loading ~ 0.090 mg$_{Pt\text{-}cathode}$/cm$^2$ under $H_2/O_2$ conditions. PtNi/PECC demonstrates an ORR MA of 0.93 ± 0.06 A/mg$_{Pt}$, exceeding the DOE's 2020 target (0.44 A/mg$_{Pt}$).[11] It is also 63% higher than the MA of PtNi/CNT+I (0.57 ± 0.07 A/mg$_{Pt}$) and 90% higher than the MA of c-PtNi/C+I (0.49 ± 0.05 A/mg$_{Pt}$) (Fig. 3a), demonstrating its potential for high system efficiency.[45] Given similar electrochemical active surface area (ECSA) (Extended Data Fig. 3a), PtNi NCs in PtNi/PECC exhibit a specific activity (SA) at 2.07 ± 0.08 mA/cm$^2_{Pt}$, 64% higher than that in PtNi/CNT+I (1.26 ± 0.08 mA/cm$^2_{Pt}$) and 120% higher than that in c-PtNi/C+I (0.94 ± 0.13 mA/cm$^2_{Pt}$) (Extended Data Fig. 3b). **This catalytic activity enhancement can be attributed to alleviating ionomer poisoning in PtNi/PECC due to the PECC design.** Indeed, the CO displacement experiment (Method 4.4) confirmed that the PtNi NCs in PtNi/PECC show the lowest surface coverage of 4.2% at 0.4 V (Extended Data Fig. 3c), much lower than that in PtNi/CNT+I (11.1%) and c-PtNi/C+I (12.7%) (Extended Data Fig. 3d), representing the lowest values among state-of-the-art studies (Supplementary Table 4).

The PEMFCs were further tested under $H_2$/air conditions at 80 °C, 100% RH, 150 kPa$_{Abs}$. The PtNi/PECC delivers 0.30 A/cm$^2$ at 0.8 V that reaches the DOE ultimate target for light-duty vehicles (0.3 A/cm$^2$ at 0.8 V),[11] which is **150% higher than that of c-PtNi/C+I (0.12 A/cm$^2$) and 88% higher than that of PtNi/CNT+I** (0.16 A/cm$^2$) (Fig. 3b). In the high-current-density (HCD) region, where the performance determines the startup and acceleration for vehicles[24] and the volumetric power density,[2,10] the **PtNi/PECC achieves a peak power of 1.51 W/cm$^2$**, **34% higher than that of c-PtNi/C+I (**1.13 W/cm$^2$**) and 57% higher than that of PtNi/CNT+I** (0.96 W/cm$^2$). **The PtNi/PECC also achieves a rated power of 1.01W/cm$^2$ (at 0.67 V), which is 40% higher than that of c-PtNi/C+I** (0.72 W/cm$^2$) **and 42% higher than that of PtNi/CNT+I** (0.71 W/cm$^2$).

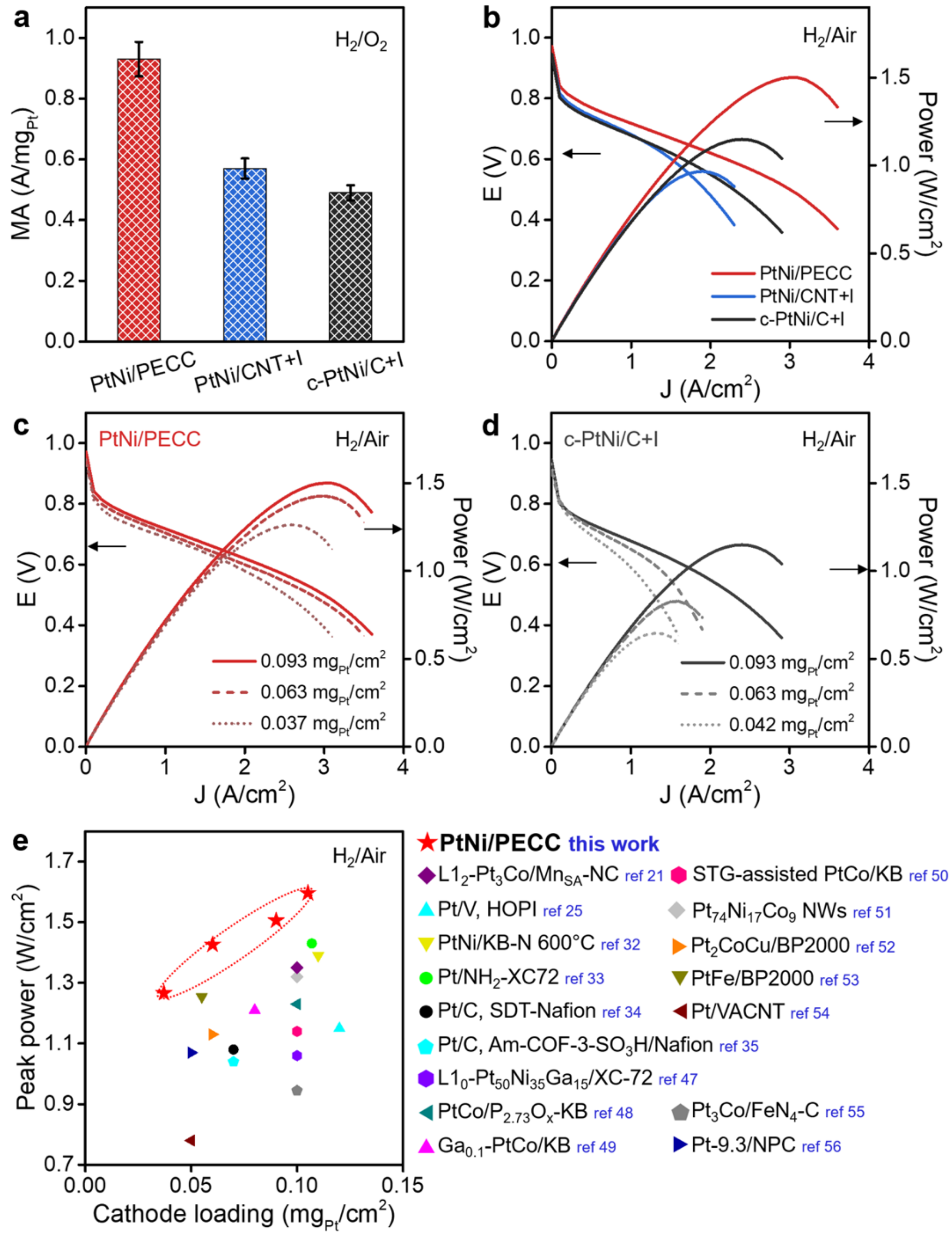

**Fig. 3. | The performance of MEA with PECC CCL (PtNi/PECC) vs. conventional CCLs fabricated with PtNi/CNT + ionomer (PtNi/CNT+I) and commercial PtNi/C + ionomer (c-PtNi/C+I). a**, The mass activity (MA) of PtNi nanocatalysts (NCs) in three different CCL measured at $H_2/O_2$, 80 °C, 100% RH, 150 $kPa_{Abs}$. **b**, The polarization and power curves of MEAs with different CCLs measured at $H_2$/Air, 80 °C, 100% RH, 150 $kPa_{Abs}$. The loadings are 0.093, 0.092, and 0.093 $mg_{Pt\text{-}cathode}/cm^2$ for PtNi/PECC, PtNi/CNT+I, and c-PtNi/C+I, respectively. **c**, The polarization and power curves of PtNi/PECC with ultralow cathode Pt loadings. **d**, The polarization and power curves of c-PtNi/C+I with ultralow cathode Pt loadings. **e**, The comparison of peak power density in this work to state-of-the-art publications. A detailed comparison is presented in Supplementary Table 7. All Pt loading ($mg_{Pt}/cm^2$) labels refer to the cathode Pt loading.

To better understand the origins of the observed power enhancement in PtNi/PECC, we deconvoluted the polarization curves to quantitatively assess activity overpotential ($\eta_{activity}$), ohmic overpotential ($\eta_{ohmic}$), and mass transport overpotential ($\eta_{mass\ transport}$), respectively.[46] (Extended Data Fig. 4a-c, Supplementary Table 5, and Method 4.5). First of all, the PtNi/PECC shows the smallest $\eta_{activity}$ (Extended Data Fig. 4a) among all tested CCLs. At 0.3 A/cm$^2$, the $\eta_{activity}$ of PtNi/PECC is 366 mV, 41 mV less than that of c-PtNi/C+I (407 mV), and 29 mV less than that of PtNi/CNT+I (395 mV), consistent with its highest MA. Meanwhile, all three MEAs demonstrated similar $\eta_{ohmic}$ (Extended Data Fig. 4b) within 5% differences, suggesting membrane resistance dominates the $\eta_{ohmic}$. Further electrical analyses showed that **the PtNi/PECC** exhibits substantially lower in-plane electrical resistance than the conventional CCL, **demonstrating an electrical conductivity of 17.6 S/cm, 220% higher than that of c-PtNi/C+I (5.5 S/cm)**, corresponding to a 63.8% reduction in sheet resistance (107 vs. 294 Ω/sq). (Extended Data Fig. 5, Method 4.6 and 4.7)

Importantly, the PtNi/PECC demonstrates the smallest $\eta_{mass\ transport}$ (Extended Data Fig. 4c), highlighting the superior mass transport enabled by the PECC design. At 2.0 A/cm$^2$, the $\eta_{mass\ transport}$ of the PtNi/PECC is only 38 mV, 50% of that in the c-PtNi/C+I (76 mV). The significant enhancement in the mass transport in the PtNi/PECC is further manifested in **MEAs with ultralow Pt loadings ~0.040 mg$_{Pt\text{-}cathode}$/cm$^2$)**, wherein the mass transport is more demanding owing to the limited Pt surface area (Extended Data Fig. 1d). Notably, **the PtNi/PECC with ultralow Pt loading of only 0.063 mg$_{Pt\text{-}cathode}$/cm$^2$ and 0.037 mg$_{Pt\text{-}cathode}$/cm$^2$ delivered peak power density of 1.43 W/cm$^2$ and 1.26 W/cm$^2$ (Fig. 3c), respectively, which are 64% and 85% higher than c-PtNi/C+I with similar loadings** (i.e., 0.87 W/cm$^2$ at 0.063 mg$_{Pt\text{-}cathode}$/cm$^2$ and 0.68W/cm$^2$ at 0.042 mg$_{Pt\text{-}cathode}$/cm$^2$, Fig. 3d). Also, the PtNi/PECC demonstrated only minor performance degradation and when the Pt loading is lowered from 0.093 mg$_{Pt\text{-}cathode}$/cm$^2$, to 0.063 mg$_{Pt\text{-}cathode}$/cm$^2$ (5% decrease in power density) and 0.037 mg$_{Pt\text{-}cathode}$/cm$^2$ (17% decrease). In contrast, the low Pt-loading c-PtNi/C+I (0.063 mg$_{Pt\text{-}cathode}$/cm$^2$ and 0.042 mg$_{Pt\text{-}cathode}$/cm$^2$) observed far greater power decrease of 23% and 41% compared to 0.093 mg$_{Pt\text{-}cathode}$/cm$^2$ (Extended Data Fig. 6a,b). The analyses of cell overpotential at various Pt loadings (Extended Data Fig. 4d-f, Supplementary Table 6) confirmed that when the loading decreases from ~0.090 $mg_{Pt\text{-}cathode}$ /$cm^2$ to ~0.040 $mg_{Pt\text{-}cathode}$/$cm^2$, **the c-PtNi/C+I shows a dramatic 457% increase in $\eta_{mass\ transport}$ at 1.5 A/$cm^2$ (**from 35 mV to 195mV)**; while the $\eta_{mass\ transport}$ of PtNi/PECC remains unchanged** (from 19 mV to 19 mV, Extended Data Fig. 4f). At even higher current density of 2.5 A/$cm^2$, which is unreachable for the c-PtNi/C+I with low Pt loadings, the PtNi/PECC still demonstrates satisfactory MEA performance owing to its low $\eta_{mass\ transport}$.

The excellent mass transport properties of PECC CCL endow the PtNi/PECC MEA with the highest peak power among the state-of-the-art MEA performance tested at similar conditions (Fig. 3e).[21,25,32-35,47-56] At a loading of 0.125 $mg_{Pt\text{-}total}$/$cm^2$, PtNi/PECC achieved a high power of 1.60 W/$cm^2$ (Extended Data Fig. 6c), **yielding a projected volumetric power density of 8.91 kW/L that surpasses the 2030 target of 6.0 kW/L and nearly reaches the 2040 target of 9.0 kW/L**[2] (Method 4.8). This represents a significant leap towards the cost-effective commercialization of the PEMFCs, and its impact can be further enhanced through integration with complementary advances such as GDL engineering.[57] **Furthermore, the PtNi/PECC with ultralow loading 0.037 $mg_{Pt\text{-}cathode}$/$cm^2$ delivers unprecedented Pt utilization 34.1 $kW_{Peak}$/$g_{Pt\text{-}cathode}$ and 22.2 $kW_{Rated}$/$g_{Pt\text{-}cathode}$. These values represent the highest ever reported MEA power performance** (**Supplementary Table 7), surpassing the stretch target for 20 kW/$g_{Pt\text{-}cathode}$ for the first time. More importantly, the PECC showed a substantially improved $H_2$ fuel utilization efficiency, delivering 19% less $H_2$ consumption compared to conventional CCL. (**Extended Data Fig. 6d, Method 4.9).

The stability of PtNi/PECC with ~0.090 $mg_{Pt\text{-}cathode}$/$cm^2$ was evaluated using a 30,000-cycle square-wave accelerated stress test (AST). After AST, PtNi/PECC retains a MA of 0.60 A/$mg_{Pt}$, five times higher than that of c-PtNi/C+I (0.12 A/$mg_{Pt}$, Extended Data Fig. 6e). The MA retention reaches 65%, significantly exceeding that of c-PtNi/C+I (25%) and meeting the DOE target (60%). The post-AST peak power of PtNi/PECC is 0.82 W/$cm^2$, representing a 32% increase over c-PtNi/C+I (Extended Data Fig. 6f). The retained activity and device performance indicate that the PECC architecture preserves its coupled proton–electron transport during prolonged operation, sustaining its performance advantage and highlighting its practical viability for durable PEMFCs and other electrochemical devices.

**Reactant transport in PtNi/PECC**

The oxygen transport resistance of different MEAs were examined (Extended Data Fig. 7a-c, Method 4.10).[58] As expected, the $R_{total, O2}$ of PtNi/PECC is the lowest among all MEAs. Specifically, the $R_{total, O2}$ of PtNi/PECC is 76.1 ± 1.4 s/m, which is 34 % lower than that of the PtNi/CNT+I (114.6 ± 3.7 s/m) and 24% lower than that of the c-PtNi/C+I (99.6 ± 1.6 s/m) with ~0.090 $mg_{Pt\text{-}cathode}/cm^2$ at 150 $kPa_{Abs}$, agreeing with the low $\eta_{mass\ transport}$ of PtNi/PECC.

The $R_{total, O2}$ was further decomposed (dotted line in Fig. 4a and Supplementary Table 8) to Fickian component ($R_F$, pressure-dependent) and non-Fickian component ($R_{NF}$, pressure-independent),[34,35] which reveals that the major difference in $R_{total, O2}$ among the MEAs originates from the difference in $R_{NF, O2}$ (Fig. 4b). **The record-low $R_{NF, O2}$ of PtNi/PECC (1.1 ± 1.8 s/m) is merely 4.7% of that of the PtNi/CNT+I (23.6 ± 5.8 s/m) and 6.4% of that of the c-PtNi/C+I (17.1 ± 3.9 s/m)**. Moreover, a very limited increase in $R_{NF, O2}$ is observed in PtNi/PECC MEAs when the loading decreases from 0.093 $mg_{Pt\text{-}cathode}/cm^2$ ($R_{NF, O2}$ = 1.1 ± 1.8 s/m), to 0.063 $mg_{Pt\text{-}cathode}/cm^2$ ($R_{NF, O2}$ = 1.4 ± 1.6 s/m), and to 0.037 $mg_{Pt\text{-}cathode}/cm^2$ ($R_{NF, O2}$ =2.1 ± 1.8 s/m) (Fig. 4c, Extended Data Fig. 7d-f, and Supplementary Table 9), agreeing with the consistently low $\eta_{mass\ transport}$ observed in PtNi/PECC CCLs at ultralow loadings (Extended Data Fig. 4f). By contrast, c-PtNi/C+I CCL shows a sharp increase in $R_{NF}$ from $R_{NF, O2}$ = 17.1 ± 3.9 s/m at 0.093 $mg_{Pt\text{-}cathode}/cm^2$, to $R_{NF, O2}$ = 28.6 ± 5.4 s/m at 0.063 $mg_{Pt\text{-}cathode}/cm^2$, and to $R_{NF, O2}$ =41.3 ± 6.1 s/m at 0.042 $mg_{Pt\text{-}cathode}/cm^2$ (Fig. 4c, Extended Data Fig. 7g-i), consistent with previous studies on $R_{NF, O2}$ sensitivity to Pt loadings in ionomer-based conventional CCLs.[9,59,60]

The $R_{Pt\text{-}local, O2}$, the roughness factor normalized Pt local oxygen transport resistance, was used to analyze local oxygen transport.[59,60] Significantly, **the $R_{Pt\text{-}local, O2}$ of PtNi/PECC is only 28.0 ± 2.2 s/m, which is just ~3% of that of c-PtNi/C+I (847.9 ± 112.9 s/m)** (Fig. 4d, Supplementary Table 10), owing to the absence of ionomer in the PECC. This greatly reduced $R_{Pt\text{-}local, O2}$ preserves the ultralow $R_{NF, O2}$ at ultralow loadings (~0.060 $mg_{Pt\text{-}cathode}/cm^2$ and ~0.040 $mg_{Pt\text{-}cathode}/cm^2$) with limited roughness factor, **delivering an $R_{NF, O2}$ that is an order of magnitude lower than the current state-of-the-art values** (Fig. 4e, Supplementary Table 11).[25,33-35,61-64] The study by a multi-physics-based 2D PEMFC MEA model (Method 5, Supplementary Table 12)[65,66] demonstrates that at a catalyst loading of 0.090 $mg_{Pt\text{-}cathode}/cm^2$, conventional CCLs with ionomer-film-covered catalyst show $R_{NF, O2}$ values between 33.9 and 28.9 s/m, about ten times greater than the predicted 1.1–0.8 s/m for water-film-covered catalysts in the PECC (Fig. 4f). These findings closely match experimental data (Fig. 4b), confirming the superior oxygen transport enabled by the PECC architecture. Additionally, the simulations indicate that simply thinning the ionomer layer on the catalyst has minimal impact on reducing $R_{NF}$: decreasing the ionomer thickness from 10 nm to 0.5 nm only lowers $R_{NF, O2}$ by 18% (Fig. 4f). This aligns with prior studies showing that even a 0.5 nm ionomer coating on Pt accounts for most of the local oxygen transport resistance (Extended Data Fig. 1c),[9,25,67,68] highlighting the importance of removing the Pt–ionomer interface in CCLs to significantly enhance oxygen transport.

This PECC design hence resolves a key limitation (i.e., $R_{NF, O2}$) in PEMFC power performance, especially in those with ultralow Pt loadings, paving the way for cost-effective high-power performance PEMFCs. Importantly, the PECC CL design has also been successfully implemented in the catalyst layers of EHPs (Extended Data Fig. 8-9, Method 4.11) and **showed that PtNi/PECC exhibits much lower $R_{NF, H2}$ (20.7 s/m), decreased by 92% from 248 s/m in c-PtNi/C+I** (Fig. 4g), thus requiring substantially lower cell voltage to deliver the same hydrogen production. This suggests the potential of PECC architecture to broadly mitigate mass transport bottlenecks in various electrochemical devices.

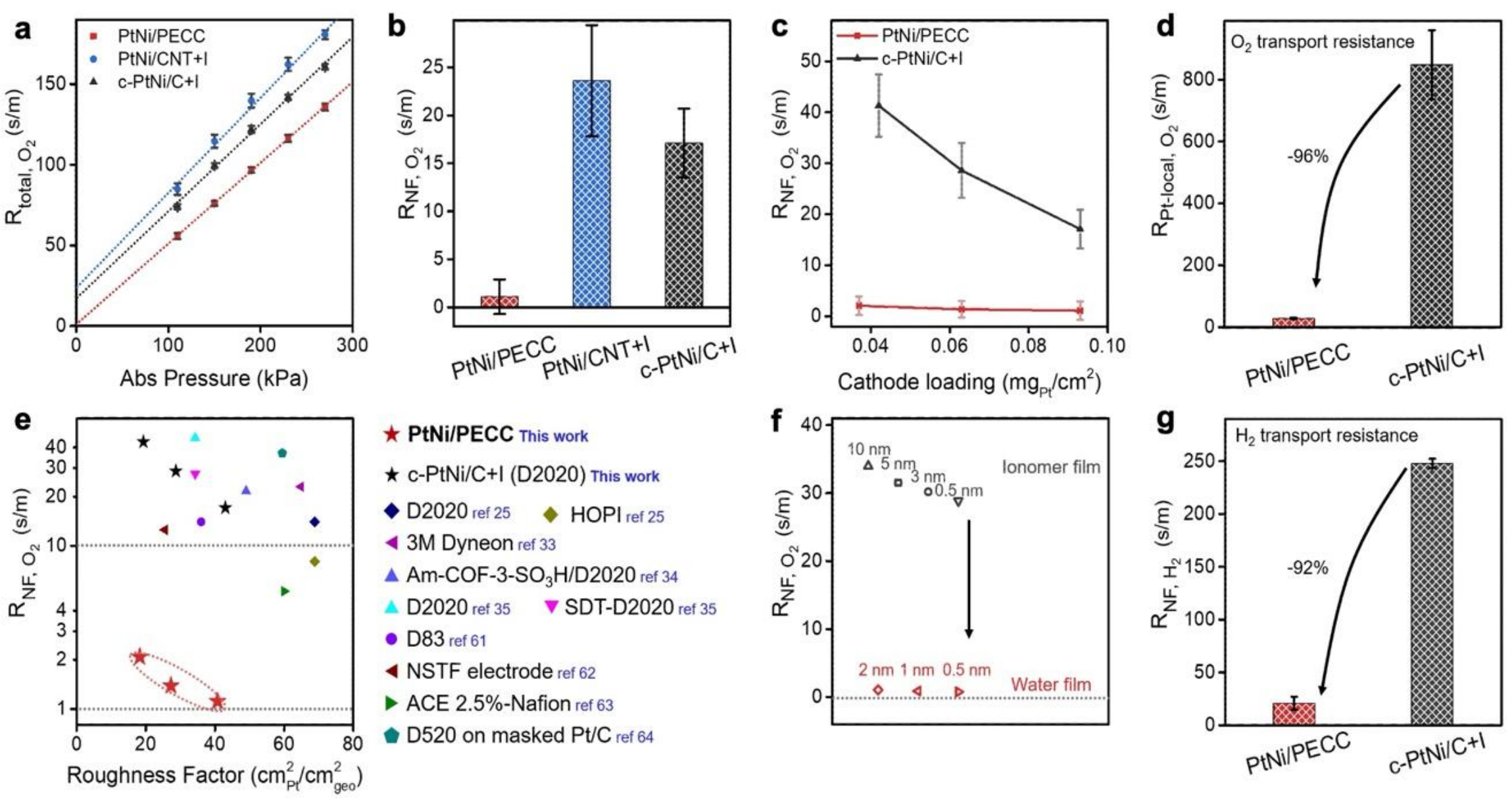

**Fig. 4. | The mass transport in PtNi/PECC, PtNi/CNT+I, and c-PtNi/C+I. a**, The total oxygen transport resistance ($R_{total, O2}$) of different CCLs. **b**, The comparison of non-Fickian oxygen transport resistance ($R_{NF, O2}$) of different CCLs. **c**, The $R_{NF, O2}$ of different CCLs at decreased cathode Pt loading. **d**, The Pt local oxygen transport resistance ($R_{Pt\text{-}local, O2}$) of different CCLs. **e**, The comparison of $R_{NF, O2}$ vs roughness factor obtained in this work to the state-of-the-art values measured under similar conditions. A detailed comparison is provided in Supplementary Table 11. **f**, Multi-physics-based continuum simulation on $R_{NF, O2}$ of PtNi NCs covered by ionomer or water film with various film thicknesses. **g**, The non-Fickian hydrogen transport resistance ($R_{NF, H2}$) measured in an EHP with conventional or PECC CL as anode for HOR (Pt loading 0.01 $mg_{Pt}/cm^2$).

## Proton transport in PtNi/PECC CCL

The excellent cell-level performances enabled by PtNi/PECC also suggest sufficient proton transport to sustain the ORR without the ionomer. Indeed, Electrochemical Impedance Spectroscopy (EIS) proton transport measurement of PtNi/PECC and c-PtNi/C+I showed that the proton conductivity of the PtNi/PECC (1.19 S/m) is similar to that of c-PtNi/C+I (1.14 S/m) (Fig. 5a-b, Supplementary Table 13, Method 4.12). Correspondingly, the proton transport resistance of PtNi/PECC was 45.4 mΩ·$cm^2$, 16% lower than that in c-PtNi/C+I (54.3 mΩ·$cm^2$) (Fig. 5c).

Molecular dynamics (MD) simulations were performed to elucidate the proton transport mechanism in the PECC (Method 6, Extended Data Fig. 10a-c), which was compared with that in Nafion under a similar hydration level (Extended Data Fig. 10d). To mimic the PECC support, we constructed a model of a

sulfonate–grafted graphene surface under high humidity (λ = 15), populated with benzenesulfonate groups, water, oxygen, and hydronium ions (Extended Data Fig. 10b-c). A representative side view (Fig. 5d) and the corresponding species density profile (Fig. 5e, Extended Data Fig. 10e,f) show the strong affinity of sulfonate groups for water and hydronium, forming a five-angstrom-thick water-hydronium sheath (termed the "IonoSkin") spanning 4 Å to 9 Å above the surface, which hosts all hydronium ions of the system. The top view further demonstrates the long-range connectivity of the IonoSkin, a continuity that underpins proton transport in the PECC CLs (Fig. 5f).

The IonoSkin's surface-guided water and hydronium distribution may suggest an anisotropic transport mode for protons on PECC (Fig. 5d,f). The x, y, and z components of the effective vehicular diffusivity ($D_x$, $D_y$, and $D_z$) were deconvoluted to demonstrate the influence of the water–hydronium network (WHN) boundaries, which confine proton transport within the connected domains (Method 6.6). The in-plane components ($D_x$, $D_y$) of the IonoSkin are five orders of magnitude higher than its out-of-plane component ($D_z$), confirming a surface-guided 2D transport mode (Fig. 5g). By contrast, Nafion shows comparable $D_x$, $D_y$, and $D_z$, consistent with its three-dimensional (3D) isotropic transport despite minor differences due to finite sampling (Fig. 5g). Hence, we applied a 2D diffusion model for the simulation of the vehicular and hopping proton diffusivity in IonoSkin, while adopting a 3D diffusion model for Nafion (Supplementary Table 14).

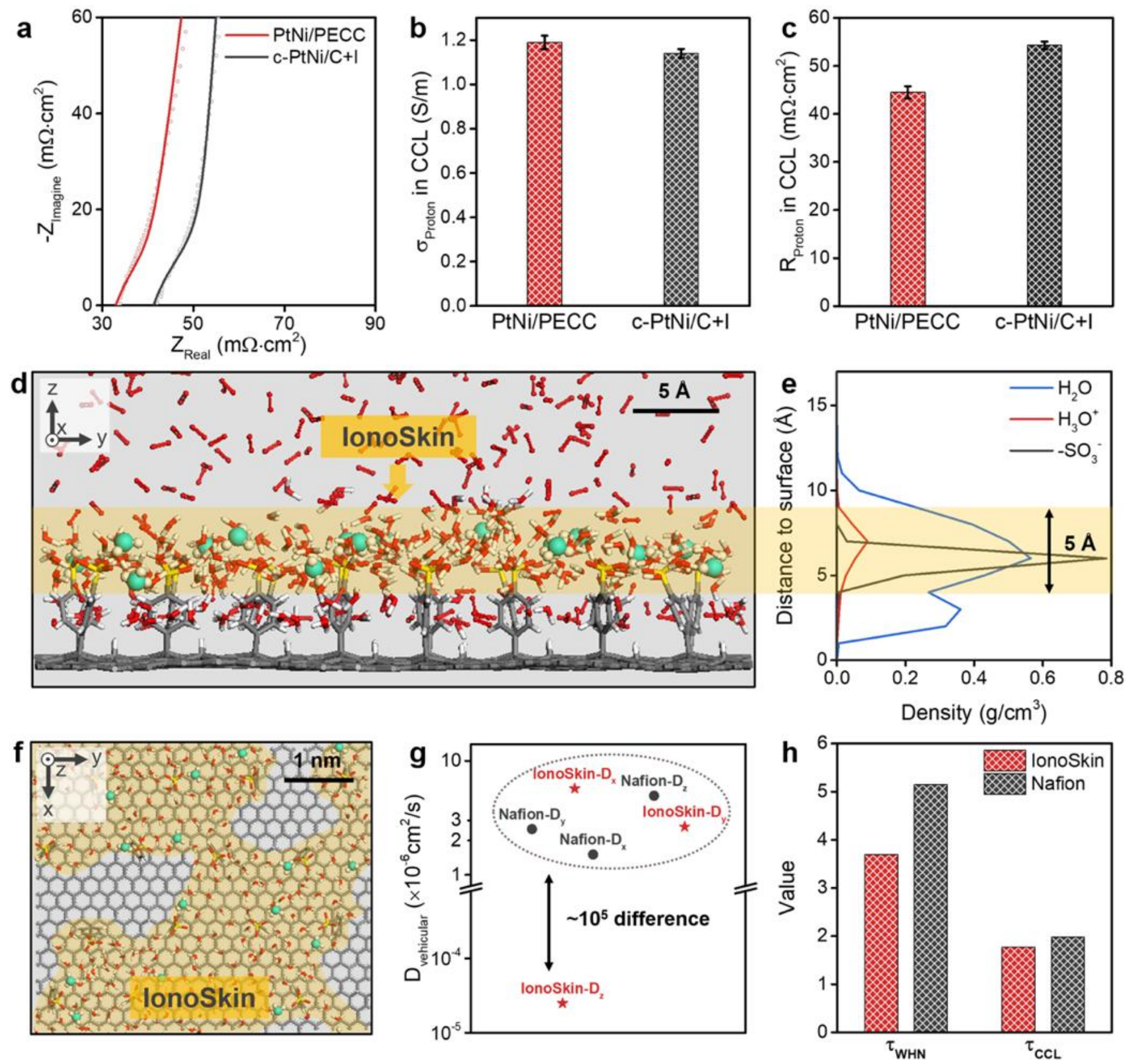

**Fig. 5. | The proton transport in PtNi/PECC. a-c,** Proton transport resistance of different CCLs measured by Electrochemical Impedance Spectroscopy (EIS) under $H_2/N_2$: **a**, Nyquist plot of PtNi/PECC and c-PtNi/C+I; **b**, The fitted proton conductivity of different CCLs. **c**, The fitted active-area-normalized proton transport resistance of different CCL. **d-f**, A typical system configuration in which 16 hydroniums, 224 $H_2O$, and 413 $O_2$ molecules are present for the proton transport simulation at $\lambda$ =15 and $O_2$ pressure=200 atm. Red = oxygen; Grey = carbon; White = hydrogen; Yellow = sulfur. The oxygen of hydronium was highlighted in cyan. The proton-conductive sheath (IonoSkin) was highlighted in translucent yellow: **d**, The side view of the configuration (x direction comes out toward the viewer); **e**, The species density profile based on the distance from the surface over the last 20 ns in equilibrium state, corresponding to (D); **f**, The top view of the configuration (z direction comes out towards the viewer). The oxygen molecules are omitted for better clarity. **g**, Simulated effective vehicular diffusivity ($D_{vehicular}$) in x, y, and z directions of IonoSkin and Nafion. **h**, Transport tortuosity factors of water-hydronium network (WHN) within IonoSkin and Nafion ($\tau_{WHN}$) and the tortuosity factor of IonoSkin and Nafion distribution in the CCL ($\tau_{CCL}$).

To further understand the proton conductance in IonoSkin in PECC CL and in Nafion in conventional CL, we further analyze tortuosity factors at the molecular level and CL level.

At the molecular level, the tortuosity factor of proton transport can be derived from its vehicular diffusivity in the water-hydronium network (WHN), termed $\tau_{WHN}$ (Method 6.7).[69,70] The $\tau_{WHN}$ of the IonoSkin is calculated to be 3.69, 28% lower than that of the Nafion ($\tau_{WHN} = 5.15$, Fig. 5h), indicating a more continuous and less convoluted WHN in the IonoSkin. Based on the $\tau_{WHN}$, the calculated effective total proton diffusivity for IonoSkin ($D_{IonoSkin} = 3.59 \times 10^{-5}\ cm^2/s$) is 78% higher than that of Nafion ($D_{Nafion} = 2.00 \times 10^{-5}\ cm^2/s$, Supplementary Table 14, Method 6.8-6.9). Meanwhile, the proton concentration in IonoSkin is estimated to be $2.53 \times 10^{-3}\ mol/cm^3$, 93% higher than that in the Nafion ($1.31 \times 10^{-3}\ mol/cm^3$) due to its confined volume (Method 6.10). Together, the IonoSkin presents an in-plane conductivity **$\sigma_{IonoSkin} = 28.8\ S/m$, 249% higher than that of Nafion ($\sigma_{Nafion} = 8.26\ S/m$), suggesting its excellent proton conductivity** (Method 6.11).

At the cell level, the tortuosity factor (termed $\tau_{CCL}$) and the proton conductor volume fraction determines the measured proton conductivity.[29,71] The volume fractions of the ionic conductor in both PECC CCL and Nafion CCL (Method 6.12) are determined to be 7.3 vol% and 27.3 vol%, respectively. With the measured proton conductivity in PECC CCL (1.19 S/m) and Nafion CCL (1.14 S/m), we can estimate $\tau_{CCL} = 1.77$ for PECC CCL with IonoSkin, which is 11% lower than conventional Nafion-CCL ($\tau_{CCL} = 1.98$) (Fig. 5h, Supplementary Table 15, Method 6.13). This again suggests a more connected and shorter proton percolation network in PECC CCL than in Nafion CCL.

Collectively, these findings identify IonoSkin as a distinctive surface-guided, anisotropic proton sheath that enables superior proton conductivity at both molecular and device levels, while also ensuring optimal oxygen accessibility to the catalysts.

**Conclusion**

This study introduces a proton-electron coupled catalyst (PECC) architecture designed to facilitate efficient, simultaneous mass transport of electrons and protons while providing unobstructed gas reactant access to the catalysts in electrochemical systems. This unique design significantly enhances power performance, fuel efficiency, and energy efficiency across PEMFC and EHP. Investigations into $O_2$ and $H_2$ mass transport, supported by multi-physics simulations, reveal an order-of-magnitude reduction in both non-Fickian gas transport resistance and local gas transport resistance at the catalyst surface. These improvements are attributed to the ionomer-free catalyst environment. Additionally, the PECC demonstrates superior proton conductance compared to conventional ionomer-based catalyst layers (CLs), as evidenced by PEMFC electrochemical impedance spectroscopy (EIS) analysis, HER-limited EHP performance assays, and molecular dynamics (MD) simulations. The results indicate that proton transport within the PECC's IonoSkin—characterized by an ultrathin, anisotropic water sheath confined to a sub-nanometer layer on the surface—surpasses that of conventional ionomer networks, facilitated by enhanced in-plane diffusivity and reduced tortuosity over various length scales within the CL. In practical application, the PECC CCL enabled PEMFC achieves a peak power density of 1.51 W/cm$^2$ at ~0.090 mg$_{\text{Pt-cathode}}$/cm$^2$ and maintains 1.26 W/cm$^2$ even at the low loading of 0.037 mg$_{\text{Pt-cathode}}$/cm$^2$, attaining a superior Pt utilization rate of 22.2 kW$_{\text{Rated}}$/g$_{\text{Pt-cathode}}$. Furthermore, in EHPs, PECC reduced cell voltage by approximately 51% and 30% under HOR- and HER-limited conditions, respectively. These findings highlight the versatility and broad applicability of the PECC architecture beyond PEMFCs, offering promising solutions to address mass-transport limitations in advanced electrochemical technologies.

## Reference:

1 Clarke, L. *et al. Energy Systems* in *Climate Change 2022: Mitigation of Climate Change. Contribution of Working Group III to the Sixth Assessment Report of the Intergovernmental Panel on Climate Change* (eds P. R. Shukla *et al.*) Ch. 6, 613-746 (Cambridge University Press, 2022).https://doi.org/10.1017/9781009157926.008

2 Jiao, K. *et al.* Designing the next generation of proton-exchange membrane fuel cells. *Nature* **595**, 361-369, doi:10.1038/s41586-021-03482-7 (2021).

3 Maier, M. *et al.* Mass transport in PEM water electrolysers: A review. *Int. J. Hydrogen Energy* **47**, 30-56, doi:10.1016/j.ijhydene.2021.10.013 (2022).

4 Tan, Y. C., Lee, K. B., Song, H. & Oh, J. Modulating Local CO2 Concentration as a General Strategy for Enhancing C−C Coupling in CO2 Electroreduction. *Joule* **4**, 1104-1120, doi:10.1016/j.joule.2020.03.013 (2020).

5 Cetinbas, F. C., Ahluwalia, R. K., Kariuki, N. N. & Myers, D. J. Agglomerates in Polymer Electrolyte Fuel Cell Electrodes: Part I. Structural Characterization. *J. Electrochem. Soc.* **165**, F1051-F1058, doi:10.1149/2.0571813jes (2018).

6 Girod, R., Lazaridis, T., Gasteiger, H. A. & Tileli, V. Three-dimensional nanoimaging of fuel cell catalyst layers. *Nat. Catal.* **6**, 383-391, doi:10.1038/s41929-023-00947-y (2023).

7 Schuler, T. *et al.* Fuel-Cell Catalyst-Layer Resistance via Hydrogen Limiting-Current Measurements. *J. Electrochem. Soc.* **166**, F3020-F3031, doi:10.1149/2.0031907jes (2019).

8 Möller, T. *et al.* The product selectivity zones in gas diffusion electrodes during the electrocatalytic reduction of CO2. *Energy Environ. Sci.* **14**, 5995-6006, doi:10.1039/d1ee01696b (2021).

9 Kongkanand, A. & Mathias, M. F. The Priority and Challenge of High-Power Performance of Low-Platinum Proton-Exchange Membrane Fuel Cells. *J. Phys. Chem. Lett.* **7**, 1127-1137, doi:10.1021/acs.jpclett.6b00216 (2016).

10 Fan, L. *et al.* Towards ultralow platinum loading proton exchange membrane fuel cells. *Energy Environ. Sci.* **16**, 1466-1479, doi:10.1039/D2EE03169H (2023).

11 U.S. Department of Energy. Fuel Cell Technologies Office Multi-Year Research, Development, and Demonstration Plan: Section 3.4 Fuel Cells. (U.S. Department of Energy, 2017). Available at https://www.energy.gov/sites/prod/files/2017/05/f34/fcto_myrdd_fuel_cells.pdf.

12 Li, M. *et al.* Ultrafine jagged platinum nanowires enable ultrahigh mass activity for the oxygen reduction reaction. *Science* **354**, 1414-1419, doi:doi:10.1126/science.aaf9050 (2016).

13 Chong, L. *et al.* Ultralow-loading platinum-cobalt fuel cell catalysts derived from imidazolate frameworks. *Science* **362**, 1276-1281, doi:doi:10.1126/science.aau0630 (2018).

14 Tian, X. *et al.* Engineering bunched Pt-Ni alloy nanocages for efficient oxygen reduction in practical fuel cells. *Science* **366**, 850-856, doi:doi:10.1126/science.aaw7493 (2019).

15 Huang, X. *et al.* High-performance transition metal-doped Pt3Ni octahedra for oxygen reduction reaction. *Science* **348**, 1230-1234, doi:doi:10.1126/science.aaa8765 (2015).

16 Yang, C.-L. *et al.* Sulfur-anchoring synthesis of platinum intermetallic nanoparticle catalysts for fuel cells. *Science* **374**, 459-464, doi:doi:10.1126/science.abj9980 (2021).

17 Escudero-Escribano, M. *et al.* Tuning the activity of Pt alloy electrocatalysts by means of the lanthanide contraction. *Science* **352**, 73-76, doi:doi:10.1126/science.aad8892 (2016).

18 Chen, C. *et al.* Highly Crystalline Multimetallic Nanoframes with Three-Dimensional Electrocatalytic Surfaces. *Science* **343**, 1339-1343, doi:doi:10.1126/science.1249061 (2014).

19 Zhang, L. *et al.* Platinum-based nanocages with subnanometer-thick walls and well-defined, controllable facets. *Science* **349**, 412-416, doi:doi:10.1126/science.aab0801 (2015).

20 Li, J. *et al.* Hard-Magnet L10-CoPt Nanoparticles Advance Fuel Cell Catalysis. *Joule* **3**, 124-135, doi:10.1016/j.joule.2018.09.016 (2019).

21 Zeng, Y. *et al.* Regulating Catalytic Properties and Thermal Stability of Pt and PtCo Intermetallic Fuel-Cell Catalysts via Strong Coupling Effects between Single-Metal Site-Rich Carbon and Pt. *J. Am. Chem. Soc.* **145**, 17643-17655, doi:10.1021/jacs.3c03345 (2023).

22 Cheng, H. *et al.* Subsize Pt-based intermetallic compound enables long-term cyclic mass activity for fuel-cell oxygen reduction. *Proc. Natl. Acad. Sci. U. S. A.* **118**, e2104026118, doi:doi:10.1073/pnas.2104026118 (2021).

23 Huang, L., Zheng, C. Y., Shen, B. & Mirkin, C. A. High-Index-Facet Metal-Alloy Nanoparticles as Fuel Cell Electrocatalysts. *Adv. Mater.* **32**, 2002849, doi:https://doi.org/10.1002/adma.202002849 (2020).

24 Kodama, K., Nagai, T., Kuwaki, A., Jinnouchi, R. & Morimoto, Y. Challenges in applying highly active Pt-based nanostructured catalysts for oxygen reduction reactions to fuel cell vehicles. *Nat. Nanotechnol.* **16**, 140-147, doi:10.1038/s41565-020-00824-w (2021).

25 Jinnouchi, R. *et al.* The role of oxygen-permeable ionomer for polymer electrolyte fuel cells. *Nat. Commun.* **12**, 4956, doi:10.1038/s41467-021-25301-3 (2021).

26 Garsany, Y. *et al.* Improving PEMFC Performance Using Short-Side-Chain Low-Equivalent-Weight PFSA Ionomer in the Cathode Catalyst Layer. *J. Electrochem. Soc.* **165**, F381-F391, doi:10.1149/2.1361805jes (2018).

27 Park, Y.-C., Tokiwa, H., Kakinuma, K., Watanabe, M. & Uchida, M. Effects of carbon supports on Pt distribution, ionomer coverage and cathode performance for polymer electrolyte fuel cells. *J. Power Sources* **315**, 179-191, doi:10.1016/j.jpowsour.2016.02.091 (2016).

28 Yarlagadda, V. *et al.* Boosting Fuel Cell Performance with Accessible Carbon Mesopores. *ACS Energy Lett.* **3**, 618-621, doi:10.1021/acsenergylett.8b00186 (2018).

29 Liu, Y., Ji, C., Gu, W., Jorne, J. & Gasteiger, H. A. Effects of Catalyst Carbon Support on Proton Conduction and Cathode Performance in PEM Fuel Cells. *J. Electrochem. Soc.* **158**, B614-B621, doi:10.1149/1.3562945 (2011).

30 Yarlagadda, V., Mellott, N., Kumaraguru, S. & Ramaswamy, N. Proton Transport Functionality-Enabled Carbon Support for Improved Fuel Cell Performance and Durability. ACS Appl. Mater. Interfaces 15, 55669-55678, doi:10.1021/acsami.3c11528 (2023).

31 Horst, R. J. & Forner-Cuenca, A. Low-Ionomer Polymer Electrolyte Fuel Cells Enabled by Direct Acid Grafting of Pt/C Catalysts. J. Electrochem. Soc. 173, doi:10.1149/1945-7111/ae4ff9 (2026).

32 Li, C. *et al.* Unraveling the core of fuel cell performance: engineering the ionomer/catalyst interface. *Energy Environ. Sci.* **16**, 2977-2990, doi:10.1039/d2ee03553g (2023).

33 Ott, S. *et al.* Ionomer distribution control in porous carbon-supported catalyst layers for high-power and low Pt-loaded proton exchange membrane fuel cells. *Nat. Mater.* **19**, 77-85, doi:10.1038/s41563-019-0487-0 (2020).

34 Yang, J. *et al.* Oxygen- and proton-transporting open framework ionomer for medium-temperature fuel cells. *Science* **385**, 1115-1120, doi:10.1126/science.adq2259 (2024).

35 Zhang, Q. *et al.* Covalent organic framework–based porous ionomers for high-performance fuel cells. *Science* **378**, 181-186, doi:10.1126/science.abm6304 (2022).

36 Chen, F. *et al.* Blocking the sulfonate group in Nafion to unlock platinum's activity in membrane electrode assemblies. *Nat. Catal.* **6**, 392-401, doi:10.1038/s41929-023-00949-w (2023).

37 Lee, C. *et al.* Grooved electrodes for high-power-density fuel cells. *Nat. Energy* **8**, 685-694, doi:10.1038/s41560-023-01263-2 (2023).

38 Yang, G. *et al.* Advanced Electrode Structures for Proton Exchange Membrane Fuel Cells: Current Status and Path Forward. *Electrochem. Energy Rev.* **7**, 9, doi:10.1007/s41918-023-00208-3 (2024).

39 Sinha, P. K., Gu, W., Kongkanand, A. & Thompson, E. Performance of Nano Structured Thin Film (NSTF) Electrodes under Partially-Humidified Conditions. *J. Electrochem. Soc.* **158**, B831, doi:10.1149/1.3590748 (2011).

40 Thompson, E. L. & Baker, D. Proton Conduction on Ionomer-Free Pt Surfaces. *ECS Trans.* **41**, 709, doi:10.1149/1.3635605 (2011).

41 Yoshihara, R. et al. Ionomer-free electrocatalyst using acid-grafted carbon black as a proton-conductive support. J. Power Sources 529, doi:10.1016/j.jpowsour.2022.231192 (2022).

42 Liu, Z. *et al.* Pt catalyst protected by graphene nanopockets enables lifetimes of over 200,000 h for heavy-duty fuel cell applications. *Nat. Nanotechnol.* **20**, 807-814, doi:10.1038/s41565-025-01895-3 (2025).

43 Peng, B. *et al.* Embedded oxide clusters stabilize sub-2 nm Pt nanoparticles for highly durable fuel cells. *Nat. Catal.* **7**, 818-828, doi:10.1038/s41929-024-01180-x (2024).

44 Zhao, Z. *et al.* Graphene-nanopocket-encaged PtCo nanocatalysts for highly durable fuel cell operation under demanding ultralow-Pt-loading conditions. *Nat. Nanotechnol.* **17**, 968-975, doi:10.1038/s41565-022-01170-9 (2022).

45 Lohse-Busch, H. *et al.* Automotive fuel cell stack and system efficiency and fuel consumption based on vehicle testing on a chassis dynamometer at minus 18 °C to positive 35 °C temperatures. *Int. J. Hydrogen Energy* **45**, 861-872, doi:10.1016/j.ijhydene.2019.10.150 (2020).

46 Gerhardt, M. R. *et al.* Method—Practices and Pitfalls in Voltage Breakdown Analysis of Electrochemical Energy-Conversion Systems. *J. Electrochem. Soc.* **168**, 074503, doi:10.1149/1945-7111/abf061 (2021).

47 Liang, J. *et al.* Metal bond strength regulation enables large-scale synthesis of intermetallic nanocrystals for practical fuel cells. *Nat. Mater.* **23**, 1259-1267, doi:10.1038/s41563-024-01901-4 (2024).

48 Hu, S.-N. *et al.* A P–O functional group anchoring Pt–Co electrocatalyst for high-durability PEMFCs. *Energy Environ. Sci.* **17**, 3099-3111, doi:10.1039/d3ee04503j (2024).

49 Shao, R. Y. *et al.* Promoting ordering degree of intermetallic fuel cell catalysts by low-melting-point metal doping. *Nat. Commun.* **14**, 5896, doi:10.1038/s41467-023-41590-2 (2023).

50 Song, T.-W. *et al.* Small molecule-assisted synthesis of carbon supported platinum intermetallic fuel cell catalysts. *Nat. Commun.* **13**, 6521 (2022).

51 Liang, J. *et al.* Gas-balancing adsorption strategy towards noble-metal-based nanowire electrocatalysts. *Nat. Catal.* **7**, 719-732, doi:10.1038/s41929-024-01167-8 (2024).

52 Yin, P. *et al.* Machine-learning-accelerated design of high-performance platinum intermetallic nanoparticle fuel cell catalysts. *Nat. Commun.* **15**, 415, doi:10.1038/s41467-023-44674-1 (2024).

53 Zeng, W. J. *et al.* Phase diagrams guide synthesis of highly ordered intermetallic electrocatalysts: separating alloying and ordering stages. *Nat. Commun.* **13**, 7654, doi:10.1038/s41467-022-35457-1 (2022).

54 Meng, Q. H. *et al.* High-performance proton exchange membrane fuel cell with ultra-low loading Pt on vertically aligned carbon nanotubes as integrated catalyst layer. *J. Energy Chem.* **71**, 497-506, doi:10.1016/j.jechem.2022.03.018 (2022).

55 Qiao, Z. *et al.* Atomically dispersed single iron sites for promoting Pt and Pt3Co fuel cell catalysts: performance and durability improvements. *Energy Environ. Sci.* **14**, 4948-4960, doi:10.1039/d1ee01675j (2021).

56 Zhu, S. *et al.* Stabilized Pt Cluster-Based Catalysts Used as Low-Loading Cathode in Proton-Exchange Membrane Fuel Cells. *ACS Energy Lett.* **5**, 3021-3028, doi:10.1021/acsenergylett.0c01748 (2020).

57 Zhang, G. *et al.* Structure Design for Ultrahigh Power Density Proton Exchange Membrane Fuel Cell. *Small Methods* **7**, e2201537, doi:10.1002/smtd.202201537 (2023).

58 Baker, D. R., Caulk, D. A., Neyerlin, K. C. & Murphy, M. W. Measurement of Oxygen Transport Resistance in PEM Fuel Cells by Limiting Current Methods. *J. Electrochem. Soc.* **156**, B991, doi:10.1149/1.3152226 (2009).

59 Greszler, T. A., Caulk, D. & Sinha, P. The Impact of Platinum Loading on Oxygen Transport Resistance. *J. Electrochem. Soc.* **159**, F831-F840, doi:10.1149/2.061212jes (2012).

60 Owejan, J. P., Owejan, J. E. & Gu, W. Impact of Platinum Loading and Catalyst Layer Structure on PEMFC Performance. *J. Electrochem. Soc.* **160**, F824-F833, doi:10.1149/2.072308jes (2013).

61 Zhao, Z. *et al.* Tailoring a Three-Phase Microenvironment for High-Performance Oxygen Reduction Reaction in Proton Exchange Membrane Fuel Cells. *Matter* **3**, 1774-1790, doi:10.1016/j.matt.2020.09.025 (2020).

62 Kongkanand, A. *et al.* Degradation of PEMFC Observed on NSTF Electrodes. *J. Electrochem. Soc.* **161**, F744-F753, doi:10.1149/2.074406jes (2014).

63 Hutapea, Y. A. *et al.* Reduction of oxygen transport resistance in PEFC cathode through blending a high oxygen permeable polymer. *J. Power Sources* **556**, 232500, doi:10.1016/j.jpowsour.2022.232500 (2023).

64 Doo, G. *et al.* Nano-scale control of the ionomer distribution by molecular masking of the Pt surface in PEMFCs. *J. Mater. Chem. A* **8**, 13004-13013, doi:10.1039/c9ta14002f (2020).

65 Wang, S. & Weber, A. Z. Bridging interfacial properties and cell performance: A multiscale model for proton-exchange-membrane fuel cells. *Electrochim. Acta* **541**, 147322, doi:10.1016/j.electacta.2025.147322 (2025).

66 Chowdhury, A., Darling, R. M., Radke, C. J. & Weber, A. Z. Modeling Water Uptake and Pt Utilization in High Surface Area Carbon. *ECS Trans.* **92**, 247, doi:10.1149/09208.0247ecst (2019).

67 Jinnouchi, R., Kudo, K., Kitano, N. & Morimoto, Y. Molecular Dynamics Simulations on O 2 Permeation through Nafion Ionomer on Platinum Surface. *Electrochim. Acta* **188**, 767-776, doi:10.1016/j.electacta.2015.12.031 (2016).

68 Shinozaki, K. *et al.* Investigation of gas transport resistance in fuel cell catalyst layers via hydrogen limiting current measurements of CO-covered catalyst surfaces. *J. Power Sources* **565**, 232909, doi:10.1016/j.jpowsour.2023.232909 (2023).

69 Choi, P., Jalani, N. H. & Datta, R. Thermodynamics and Proton Transport in Nafion: II. Proton Diffusion Mechanisms and Conductivity. *J. Electrochem. Soc.* **152**, E123, doi:10.1149/1.1859814 (2005).

70 Zhao, Q., Majsztrik, P. & Benziger, J. Diffusion and interfacial transport of water in Nafion. *J. Phys. Chem. B* **115**, 2717-2727, doi:10.1021/jp1112125 (2011).

71 Ramaswamy, N. & Wortman, J. Durable Fuel Cell Electrode Design via Efficient Distribution of the Acidic Ionomer. *ACS Energy Lett.* **9**, 6170-6177, doi:10.1021/acsenergylett.4c02704 (2024).

## Methods

### 1. Chemicals and Materials

Multi-walled carbon nanotube (MWCNT) with outside diameter = 8-15 nm was purchased from Cheap Tubes Inc. $HNO_3$ (70 wt%), $H_2SO_4$ (98 wt%), HCl (35-37 wt%), acetone, ethanol, and isopropanol were purchased from Fisher Scientific. $Pt(acac)_2$ was purchased from ACROS Organics. $Ni(acac)_2$ was purchased from Aldrich. Sulfanilic acid and $NaNO_2$ were purchased from Sigma-Aldrich. Commercial Pt/C (10 wt% Pt) was purchased from Alfa Aesar. Commercial $Pt_3Ni/C$ (20 wt% Pt) was purchased from Premetek. The gas diffusion layer (GDL) and polytetrafluoroethylene (PTFE) gasket were purchased from The Fuel Cell Store. The ultrapure water (18.2MΩ cm) was generated by Millipore equipment. Nafion D2020 (20 wt%) was purchased from Ion Power Inc.

### 2. Catalyst Preparation

#### 2.1 Synthesis of PtNi/CNT

2 g of MWCNT and 20 ml of $HNO_3$ (70 wt%) were added to a 250 ml flask. After stirring and sonication, 60 ml $H_2SO_4$ (98 wt%) was slowly added to the mixture. The mixture was stirred and sonicated for 30 minutes to form a uniform slurry. The mixture was then kept at 80 °C for 1 hour. After the treatment, ultrapure ice water was slowly added to the mixture, followed by centrifugation. The treated MWCNT was washed by sonication and centrifugation till the water dispersion of MWCNT reached pH=7. The treated MWCNT was then dried under vacuum at room temperature.

The synthesis of PtNi/CNT follows the reported protocol.[44] 170 mg of $Pt(acac)_2$, 60 mg of $Ni(acac)_2$, 200 mg of treated MWCNT, and 10 ml of acetone were mixed and slowly evaporated at room temperature. The sample was collected and heat-treated at 400 °C for 6 hours under an Argon environment. The product was further acid-washed with 0.2 M $H_2SO_4$ at 85 °C and kept for 16 hours. After that, the acid was removed by repeated sonication and centrifugation in water till the pH of its water dispersion reached 7. The acid-washed PtNi/CNT was then collected by centrifugation and dried under vacuum at room temperature. After drying, the obtained catalyst was annealed in an $H_2/Ar$ mixture at 180 °C for 1.5 hours to obtain the final product.

For a fair comparison, the commercial 20 wt% $Pt_3Ni/C$ was also acid-washed and annealed with the same protocol to provide a similar Ni at% in the catalyst. The acid-washed commercial sample was labeled as "c-PtNi/C"

2.2 Synthesis of PtNi/PECC

10 ml HCl (35 wt%) and 500 mg sulfanilic acid (4-aminobenzenesulfonic acid) were dissolved in 150 ml ultrapure water with an ice bath. Then, a solution of 200 mg $NaNO_2$ dispersed in 2 ml ultrapure water was slowly added to the solution with vigorous stirring. After 2 minutes, 70 mg of PtNi/CNT was then added to the mixture and stirred for 3 days. The product is collected by filtration and washed with water, 3.5 wt% HCl, and water again till the pH of the filtrate reaches pH = 7.

**3. Characterization**

X-ray Diffraction (XRD) was acquired from a Panalytical X'Pert Pro X-ray powder diffractometer using Cu Kα as an X-ray source. X-ray Photoelectron Spectroscopy (XPS) was acquired from a Kratos AXIS Ultra DLD spectrometer. The composition of metallic elements in the catalyst was analyzed with Inductively Coupled Plasma Atomic Emission Spectroscopy (ICP-AES, Shimadzu ICPE-9000).

High-resolution Transmission Electron Microscopy (HRTEM) images were taken from the FEI TITAN 80-300 TEM operated at 300 kV. The High-Angle Annular Dark-Field image (HAADF) and Energy Dispersive Spectroscopy mapping (EDS mapping) were taken under Scanning Transmission Electron Microscopy (STEM) mode of a JEOL Grand ARM300CF STEM/TEM operated at 300kV or 80 kV. Lacey carbon-film-coated copper grids were used for all TEMs/STEMs. Scanning electron microscopy (SEM) images were taken by a ZEISS Supra 40VP SEM.

**4. Proton-Exchange Membrane Fuel Cell (PEMFC) and Electrochemical Hydrogen Pump (EHP) Measurement and Analysis**

4.1 Membrane electrode assembly (MEA) fabrication

The MEA fabrication of the conventional cathode catalyst layer (CCL) with ionomer followed the established protocol.[44] The MEA has a 5 $cm^2$ active area surrounded by a PTFE gasket. The conventional cathode catalyst ink was prepared by mixing the catalyst with water, isopropanol, and ionomer solution (Nafion D2020). The ionomer to carbon weight ratio (I/C ratio) of commercial c-PtNi/C in the cathode is 0.8. The corresponding MEA was labeled as "c-PtNi/C+I". The I/C ratio of PtNi/CNT in the cathode is 0.4. The corresponding MEA was labeled as "PtNi/CNT+I". For PtNi/PECC CCL, the cathode catalyst ink was prepared by mixing PtNi/PECC with water and isopropanol, without any ionomer.

The cathode catalyst ink was then sprayed on a 12 μm membrane using the Sono-Tek ultrasonic spray system. After that, the anode ink was sprayed on the other side of the membrane. The anode catalyst ink was prepared by mixing the commercial 10 wt% Pt/C with water, isopropanol, and ionomer with I/C = 0.7. The cathode and anode Pt loading was controlled by the spraying and was confirmed by ICP-AES. The anode Pt loading was fixed to $0.02mg_{Pt}/cm^2$ in all MEAs. After spraying, the catalyst-coated membrane was dried under vacuum, followed by pressing two GDLs to form the 5-layer MEA. Then the MEA was loaded in a 5 $cm^2$ single-cell fixture and measured by a Scribner 850e fuel cell test system.

4.2 MEA conditioning protocol

The conditioning protocol was also adopted from previous papers in our group.[44] The MEA was first activated by holding the cell potential at 0.5 V under $H_2$/Air at 80 °C, 150 $kPa_{Abs}$ (Abs = absolute pressure) and 100% Relative Humidity (RH). Then the MEA was further activated at a high humidity condition with cell temperature at 60 °C and fuel temperature at 80 °C. The same conditioning protocol was used for all MEAs.

4.3 Mass activity (MA), electrochemical surface area (ECSA), polarization curve measurement, and accelerated stressed test (AST) protocol

The MA was tested at 80 °C, 150 $kPa_{Abs}$, and 100% RH with $H_2/O_2$ at 835/2000 sccm for the anode/cathode with ohmic resistance correction (iR correction) and $H_2$ crossover correction. The $H_2$ crossover is measured under $H_2/N_2$ at 200/200 sccm.

The ECSA was measured by CO stripping. Briefly, 2.0% CO (diluted by $N_2$) was introduced to the cathode of the fuel cell, then an excess amount of pure $N_2$ gas was used to remove non-adsorbed CO. After that, a cyclic voltage scan was applied to the cell, and the area of the CO stripping peak was integrated. The success of CO stripping was confirmed by the absence of any CO oxidation peak in the second cycle of cyclic voltage scanning. The ECSA was calculated by dividing the peak area of CO oxidation by the voltage scan rate and 420 μC/cm$^2_{Pt}$.

Polarization curves were measured at 80 ℃, 150 kPa$_{Abs}$, and 100 %RH with $H_2$/Air at 835/2000 sccm. The AST consisted of 30,000 square wave cycles between 0.60 V and 0.95V (3s hold at each potential, time for changing voltage is less than 0.1s). After the AST, the post-AST MA and polarization curve is measured by the same protocol mentioned above.

4.4 CO displacement measurement

The CO displacement was measured following previous literature.[72-74] Briefly, the MEA was tested at 40 ℃, 90 %RH, 101 kPa$_{Abs}$. Prior to the measurement, the cathode was cleaned by cycling the voltage from 0.05 V to 0.90 V ten times at a scan rate of 50 mV/s under $H_2$/$N_2$. Then, the MEA was stabilized for 5 minutes at the chosen potential, ranging from 0.3 V to 0.5 V. After that, 2.0% CO (diluted by $N_2$) with a flow rate of 200 sccm was introduced to the cathode while maintaining the potential, and the cell current was recorded during the process to obtain the CO displacement curve. After the introduction of CO and the cell current became stable, the cathode was purged with an excessive amount of $N_2$ with a flow rate of 1500 sccm for 15 minutes. Lastly, a CO stripping was carried out.

The ion coverage (θ) of the Pt surface was calculated using **equation (S1)**:

$$\theta = 2 * \frac{q_{dis}}{q_{stripping}} \tag{S1}$$

where the $q_{dis}$ is the charge of the CO displacement peak, $q_{stripping}$ is the charge of the CO stripping peak.

4.5 Overpotential breakdown of the polarization curve

The overpotential breakdown was based on a previous reported method[46,75] (**equation S2-S6**).

$$V_{measured} = E_{rev} - \eta_{kinetic} - \eta_{ohmic} - \eta_{mass\ transport} \tag{S2}$$

where the $V_{measured}$ is the measured voltage at a certain current density on the polarization curve, $E_{rev}$ is the reversible cell potential of hydrogen fuel cells, $\eta_{kinetic}$ is the overpotential result from the kinetics of the catalyst, $\eta_{ohmic}$ is the result of the ohmic resistance of the fuel cell. $\eta_{mass\ transport}$ is the overpotential of the mass transport.

The $E_{rev}$ is calculated by **equation S3**.[75]

$$E_{rev} = 1.23 - 0.9 \times 10^{-3}(T - 298) + 2.3\frac{RT}{4F}\log\left(p_{H_2}^2 p_{O_2}\right) \tag{S3}$$

where T is the temperature of the system. R is the gas constant ($8.314\, J \cdot mol^{-1} \cdot K^{-1}$), F is the Faraday constant ($96485\, C \cdot mol^{-1}$), $p_{H_2}$ and $p_{O_2}$ are the partial pressures of $H_2$ and $O_2$ in the system. At 80 °C, 150 kPa, $H_2$/Air condition, the $E_{rev}$ = 1.178 V.

The $\eta_{ohmic}$ is described by the ohmic law (**equation S4**).

$$\eta_{ohmic} = i\,R \tag{S4}$$

where the $i$ is the measured current density. $R$ is the ohmic resistance of the fuel cell. It was obtained from the fuel cell station during the polarization curve measurement based on a built-in current interruption method.

The $\eta_{kinetic}$ is described by the Tafel kinetic expression (**equation S5**)

$$\eta_{kinetic} = blog\frac{i}{i_0} = a + b\log i \tag{S5}$$

where the $i$ is the measured current density, $a$ and $b$ are constant that various in different MEAs. For a measured polarization curve, the low-current region ($i \leq 0.4\ A/cm^2$) $\eta_{kinetic}$ was first calculated by subtracting $\eta_{ohmic}$ and $V_{measured}$ from the $E_{rev}$ at each current density. Then, $a$ and $b$ were obtained from linear fitting of $\eta_{kinetic}$ vs. $\log i$ within this low-current region based on **equation S5**, while the coefficient of determination ($R^2$) of this fitting was examined to make sure the Tafel relation is still valid. Lastly, the $\eta_{kinetic}$ of the whole current range was estimated based on **equation S5** with fitted $a$, $b$, and the corresponding $i$.

The $\eta_{mass\ transport}$ is estimated from the residual overpotential. It was calculated based on **equation S6**.

$$\eta_{mass\ transport} = E_{rev} - \eta_{kinetic} - \eta_{ohmic} - V_{measured} \tag{S6}$$

where the $E_{rev}$, $\eta_{ohmic}$, and $\eta_{kinetic}$ were obtained from **equation S3, S4,** and **S5** respectively. The $V_{measured}$ is the measured voltage at a certain current density on the polarization curve.

4.6 CCL electrical sheet resistance measurement and electrical conductivity calculation

The Van der Pauw method was applied to measure the sheet resistance of the CCL.[76] Conditioned MEAs were used for the measurement of more practical data. First, gaskets and GDLs of these 5-layer MEAs were removed. The residual CCM was used for the test. All CCMs were dried in vacuum overnight at room temperature to rule out the influence of water. Fast drying silver paint (TED PELLA., Inc.) was applied to the corners of the cathode active area of each CCM and dried for 30 mins. The four corners of the CCL were labelled as A, B, C, and D in a clockwise sequence. In a typical measurement, a 0.1 V voltage was applied to two adjacent corners of the square CCL (e.g., $V_{AB}$ =0.1 V), while the current across the other two adjacent corners of the CCL was measured (e.g., $I_{CD}$). This measurement was applied to AB, BC, CD, and DA of the CCL in turn. Then, the following resistances were calculated:

$$R_{AB,CD} = \frac{V_{AB}}{I_{CD}}, R_{BC,DA} = \frac{V_{BC}}{I_{DA}}, R_{CD,AB} = \frac{V_{CD}}{I_{AB}}, R_{DA,BC} = \frac{V_{DA}}{I_{BC}} \quad \text{(S7)}$$

Here, $V_{AB} = V_{BC} = V_{CD} = V_{DA} = 0.1$ V.

Also,

$$R_{horizontal} = \frac{R_{AB,CD}+R_{CD,AB}}{2}, R_{vertical} = \frac{R_{BC,DA}+R_{DA,BC}}{2} \quad \text{(S8)}$$

In our cases, the difference between the $R_{horizonal}$ and $R_{vertical}$ was less than 5% for all CCMs. Thus, we applied **equation S9** to calculate the sheet resistance

$$R_s = \frac{\pi R}{ln2} = 4.53 \times R \quad \text{(S9)}$$

Here, $R_{horizonal} = R_{vertical} = R$.

The electrical conductivity ($\sigma_{electron}$) is calculated based on **equation S10**:

$$\sigma_{electron} = \frac{1}{t \times R_s} \quad \text{(S10)}$$

where t is the thickness of the CCL measured by resin-assisted cross-section SEM (see Method 4.7).

4.7 CCL thickness measurement by cross-section SEM

Resin-assisted cross-section SEM was used for CCL thickness estimation. Conditioned MEAs were used for the measurement of a thickness that is closest to the working conditions. First, PTFE gaskets and GDLs of the 5-layer MEA were removed. The residual catalyst-coated membranes (CCMs) were used for the measurement. To prepare the sample, the CCMs were embedded in the resin (Araldite 502) and polymerized in the oven at 60 °C for at least 36 h. Then the polymerized resin was polished with P220 and P600 sand paper sequentially, followed by further polishing with 300 nm and 50 nm $Al_2O_3$ powder on a rotating polishing machine to achieve a flat surface. Lastly, a thin layer of gold was sputtered on the surface of the sample before SEM to make the surface conductive. SEM images of at least six different cross-section regions were taken, and the thickness was calculated with the help of the ImageJ software.

4.8 Estimation of the volumetric power density of PEMFCs

The volumetric power density (excluding end plates) is estimated based on structural parameters of the second-generation Toyota Mirai fuel cell stack.[77] Each single cell in this stack has a total cell thickness of 1.11 mm, consisting of a 0.511 mm thick MEA (including GDL) and anode and cathode gas channels of 0.3 mm height each. [77,78]

The MEAs used in this work exhibit a GDL-included thickness of 0.32 ± 0.02 mm for both PECC CCL and conventional CCL, closely matching the MEA thickness in the Mirai stack. This similarity suggests that our MEAs could be incorporated into the Mirai-type stack geometry, thereby justifying the following volumetric power-density estimation.

The volumetric power density ($P_V$) can be estimated by **equation S11:**

$$P_V = \frac{Power_{stack}}{V_{stack}} = \frac{N_{cell} \times A_{active} \times P_{max}}{N_{cell} \times A_{cell} \times t_{cell}} \tag{S11}$$

where $Power_{stack}$ and $V_{stack}$ is the stack power and volume, respectively. $N_{cell}$ is the number of cells inside the stack, $A_{active}$ is the MEA active area in each cell, $P_{max}$ is the peak power density of the MEA, $A_{cell}$ is the total cell area, $t_{cell}$ is the overall cell thickness, which including the thickness of MEA and two gas channels (**equation S12**).

$$t_{cell} = t_{MEA} + t_{anode\ gas\ channel} + t_{cathode\ gas\ channel} \tag{S12}$$

For our estimation:$t_{MEA} = 0.32\ mm$, $P_{max} = 1.60\ W/cm^2$ for PECC CCL and $t_{MEA} = 0.32\ mm$, $P_{max} = 1.13\ W/cm^2$ for conventional CCL. The peak power density is obtained from the polarization curves. At the same time, all other parameters were adopted from the second-generation Toyota Mirai fuel cell stack: $N_{cell} = 330$, $A_{active} = 12\ cm * 28\ cm$, $A_{cell} = 16\ cm * 41\ cm$, $t_{anode\ gas\ channel} = 0.3\ mm$, $t_{cathode\ gas\ channel} = 0.3\ mm$.

Using these parameters, the estimated volumetric power density ($P_V$) is 8.91 kW/L for MEA with PECC CCL and 6.29 kW/L for MEA with conventional CCL. The number of conventional CCL aligns well with the reported 5.4 kW/L for the Mirai stack, supporting the validity of this estimation. All volumetric power density here excludes the volume contribution of end plates.

4.9 $H_2$ consumption calculation

The hydrogen consumption per electrical energy output (g $kWh^{-1}$) was calculated from the polarization curve using Faraday's law. Assuming 100% Faradaic efficiency for the anode HOR, the mass consumption rate per geometric area is given in **equation S13**

$$\dot{m}_{H_2} = \frac{J \times M_{H_2}}{2F} \quad \text{(S13)}$$

where the $\dot{m}_{H_2}$ is the $H_2$ mass consumption rate, $J$ is the current density, $M_{H_2}$ is the $H_2$ molecular weight, F is Faraday's constant. Then, the $H_2$ consumption (g/kWh) was calculated by **equation S14**

$$H_2\ comsumption = \frac{1\ kWh}{Power} \times \dot{m}_{H_2} \quad \text{(S14)}$$

The thermodynamic lower limit is calculated based on the lower heating value (LHV, 120.9 kJ/$g_{H2}$) of hydrogen by assuming 100% conversion of the hydrogen chemical energy into electrical energy:

$$Thermodynamic\ lower\ limit\ of\ H_2\ comsumption = \frac{1\ kWh}{LHV} = 29.8\ g/kWh \quad \text{(S15)}$$

4.10 Oxygen transport resistance measurement

The oxygen transport resistance followed established methods.[44,58,61] The MEA was tested at 80 °C, 85% RH. The anode was purged with $H_2$, while two kinds of mixed gases (1.0% $O_2$ in $N_2$ and 1.5% $O_2$ in $N_2$, the number is the dry $O_2$ mole ratios) were purged to the cathode, respectively. The flow rate of $H_2$/mixed gas was 1000/1000 sccm. Three voltage sweeps from 0.45 V to 0.2 V at 2 mV/s were performed, and the average limiting current ($i_{lim}$) was obtained. The total oxygen transport resistance ($R_{total}$) was calculated from **equation S16**:

$$R_{total} = \frac{nFC_{O_2}}{i_{lim}} = \frac{nF}{i_{lim}} \times \frac{P_{Abs} - P_{H_2O}}{RT} \times x_{O_2\,dry} \tag{S16}$$

where $R_{total}$ represents the total oxygen transport resistance, $n$ represents the charge transfer number, $F$ represents the Faraday constant, $C_{O_2}$ represents oxygen concentration, $P_{Abs}$ represents the absolute pressure, $P_{H_2O}$ represents the pressure of water vapor at the testing temperature and RH, $x_{O_2\,dry}$ is the dry $O_2$ ratio, and $i_{lim}$ represents the measured limit current density.

The $R_{total}$ can be further decomposed to Fickian oxygen transport resistance ($R_F$ pressure-dependent) and non-Fickian oxygen transport resistance ($R_{NF}$, pressure-independent), as is expressed by **equation S17**.

$$R_{total} = R_F + R_{NF} = a \times P_{Abs} + R_{NF} \tag{S17}$$

where $R_F$ is the Fickian oxygen transport resistance, $R_{NF}$ is the non-Fickian oxygen transport resistance, $P_{Abs}$ is the absolute pressure, $a$ is a constant that describes the linear relation between the $R_F$ and $P_{Abs}$.

To decompose the $R_F$ and $R_{NF}$ from the $R_{total}$, the limiting current measurement was performed at absolute pressures of 110, 150, 190, 230, and 270 $kPa_{abs}$. Then, a linear fitting of $R_{total}$ vs. $P_{Abs}$ was carried out to obtain $a$, $R_F$, and $R_{NF}$.

The Knudsen resistance ($R_{Knudsen}$) and Pt-local resistance ($R_{Pt-local}$) of oxygen transport can be further deconvoluted from the $R_{NF}$ by **equation S18:**

$$R_{NF} = R_{Knudsen} + \frac{R_{Pt-local}}{Roughness\ Factor} \tag{S18}$$

where $R_{Knudsen}$ is the Knudsen oxygen transport resistance, $R_{Pt-local}$ is the Pt-local oxygen transport resistance, $Roughness\ Factor$ is the total Pt surface area per unit area of the MEA, which can be calculated from **equation S19**:

$$Roughness\ Factor = ECSA \times m_{Pt} \tag{S19}$$

Here, ECSA is the electrochemical surface area obtained from previous fuel cell measurements, $m_{Pt}$ is the cathode Pt loading.

4.11 MEA setup and test protocol for HOR-limiting and HER-limiting electrochemical hydrogen pump (EHP)

For each MEA, a counter CL and a working CL were spray-coated onto the two sides of a 12 μm PEM. The counter CL used commercial 40 wt% Pt/C with an ionomer-to-carbon (I/C) ratio of 0.8 and a Pt loading of 0.25 $mg_{Pt}/cm^2$.

The ink composition of the conventional working CL is the same as the ink used for the PEMFC; commercial c-PtNi/C and Nafion D2020 (I/C = 0.8) were used as the catalyst and ionomer, respectively. For the PECC working CL, PtNi/PECC was used without any ionomer. Both working CLs had a Pt loading of 0.01 $mg_{Pt}/cm^2$. Other fabrication steps followed the protocol used for PEMFC MEAs.

For HOR-limiting EHP measurements, the counter CL served as the cathode (HER side), and the conventional or PECC working CL served as the anode (HOR side). For HER-limiting EHP measurements, the counter CL served as the anode (HOR side), and the conventional or PECC working CL served as the cathode (HER side).

For polarization (E–J) measurements in both HOR- and HER-limiting configurations, the cell was maintained at 80 °C. The anode was supplied with 500 sccm $H_2$ (100 % RH). The cathode (HER side) was initially purged with 500 sccm $N_2$ (100 % RH) for 30 min to remove residual air, then operated in a dead-end configuration (no inlet; outlet connected to a pressure regulator with 1 atm absolute pressure). Hydrogen was generated in situ by holding the cell at 1.0 A $cm^{-2}$ for 30 min, followed by multiple potentiostatic polarization scans from 0 to 5 $A/cm^2$ at 10 mV/s until a stable curve was obtained (typically within three scans). Unless otherwise noted, the third scan is reported. The identical protocol was used for PECC and conventional working CLs.

Overpotential deconvolution of the polarization curves followed the same procedure used for PEMFC measurements (**equations S2–S5**). The MEA ohmic resistance used for the ohmic overpotential deconvolution is determined by the high-frequency resistance (HFR) at 20 kHz under 80 °C, 100%RH under $H_2/N_2$ (500sccm for both sides).

For $H_2$ transport measurement of the CLs, the EHP was operated in a HOR-limiting configuration at 80 °C, 100%RH with either the conventional or PECC CL as the anode. 800 sccm $N_2$-diluted 5% $H_2$ was fed to the anode. The cathode (HER side) was purged with 1000 sccm $N_2$ to remove the produced $H_2$ and minimize the influence of the HER side on the measurement. Polarization scans were carried out to determine the HOR limiting current at 140 kPa, 180 kPa, and 220 kPa (absolute pressure). The total $H_2$ transport resistance ($R_{total, H2}$) and the non-Fickian $H_2$ transport resistance ($R_{NF, H2}$) were calculated based on **equations S16** and **S17**

4.12 Proton conductivity measurement and fitting

The proton conductivity of CCL was measured by electrochemical impedance spectroscopy (EIS) following an established method.[79] The MEA was held at 80 ℃, 100%RH, 150 $kPa_{Abs}$ under $H_2/N_2$ with a flow rate of 500/500 sccm throughout the test. Prior to the measurement, the MEA was stabilized at 100%RH for one hour for a stable humidity. Then the EIS was measured at 0.45V with 5 mV root mean square (RMS) amplitude from 20000 Hz to 1 Hz.

With the help of Zview® software, the EIS data were fitted to an equivalent circuit **Supplementary Table 13**) that models the porous electrode. The proton resistivity ($\rho_{Proton}$, a volume normalized value with a unit of $\Omega \cdot m$) can be directly obtained from the fitting. The proton conductivity of the CCL ($\sigma_{Proton}$, a volume normalized value with a unit of $S/m$) can be calculated based on **equation S20**:

$$\sigma_{Proton} = \frac{1}{\rho_{Proton}} \tag{S20}$$

where $\rho_{Proton}$ is the proton resistivity obtained from the EIS fitting.

The area-normalized proton transport resistance of the CCL ($R_{Proton}$, unit: $\Omega \cdot cm^2$) is widely used in catalyst layer studies to evaluate the resistance of a CCL while excluding the effect of the CCL geometry area. It can be calculated based on **equation S21**:

$$R_{Proton} = t \times \rho_{Proton} \tag{S21}$$

where $t$ is the thickness of the CCL obtained from cross-section SEM, $\rho_{Proton}$ is the proton resistivity obtained from the EIS fitting.

**5. Multiphysics-based continuum modeling**

The oxygen transport resistances were calculated using an in-house developed 2D MEA PEMFC model,[65] representing the 5 $cm^2$ differential cell used in experiments. Limiting currents were determined by solving the model with $H_2$/1.0%$O_2$ in $N_2$ and $H_2$/1.5%$O_2$ in $N_2$ at 0.3V under the operation conditions of 85% RH, T=80 °C, absolute pressures ranging from 110 to 250 $kPa_{Abs}$, and different Pt loadings as used in the experiments. The non-Fickian transport resistance was then calculated as introduced in the previous section. The model was solved using the finite element method via the COMSOL Multiphysics software (version 6.0).

5.1 Model summary

While many cell-level models have been developed to explain macroscopic phenomena in PEMFCs,[80] few have focused on linking interfacial effects to cell performance, which have been demonstrated to be critical to local transport resistance, particularly at low catalyst loading in recent studies.[25,67] We use a multiscale physics-based 2D MEA model framework, which correlates the interfacial transport with cell performance[65,81] to account for transport resistance associates with interfaces, allowing differentiation between catalyst-ionomer and catalyst-water interfaces.

5.2 Model transport resistance at the interface

The 2D MEA model describes multicomponent gas diffusion, liquid and gas convection, proton and water transport, electron transport, and heat transport.[65,82] Particularly, ORR current is described using a Tafel equation combined with an agglomerate model to account for $O_2$ transport loss within the agglomerates that are formed during the fabrication process for a conventional ionomer-filled electrode. In this work, to directly compare the transport loss at catalyst-water and catalyst-ionomer interfaces involved in the ionomer-free PtNi/PECC electrode and the conventional ionomer-covered catalyst electrode, respectively, we removed the agglomerate model and modified the Tafel equation to include the interfacial transport resistance. **Equation S22** describes the ORR current density per volume of the catalyst layer after accounting for the local transport resistance,

$$i_{CL} = \frac{4F\frac{P_{O_2}}{H_{O_2,b}}S_{Pt}}{R_{int}+\frac{H_i}{H_{O_2,b}k_r}} \tag{S22}$$

where $P_{O_2}$ is the spatial oxygen partial pressure in CCL solved in the model, $H_{O_2,b}$ and $D_{O_2,b}$ are respectively the Henry's constant and diffusion coefficient of oxygen in the electrolyte bulk. $k_r$ is the intrinsic ORR rate constant described by Tafel kinetics assuming a first-order reaction,

$$k_r = \frac{i_{ref}}{4FC_{ref}} exp\left(-\frac{\alpha F}{RT}\left(\phi_{e^-} - \phi_{H^+} - E_{ref}\right)\right) \quad \text{(S23)}$$

where each term has its conventional definition. $S_{Pt}$ is the Pt area per volume of CL that can be calculated based on ECSA, Pt loading, and CCL thickness, which are all experimental variables,

$$S_{Pt} = ECSA \times \frac{Pt\ loading}{CL\ thickness} \quad \text{(S24)}$$

$R_{int}$ represents the intrinsic interfacial transport resistance from the electrolyte film on the catalyst surface. Molecular dynamics (MD) simulations suggest that both the water film and the ionomer film on the Pt surface can be separated into two regions: a few dense electrolyte molecule layers adjacent to the Pt surface, with a thickness of $l_i$ and the above bulk region with a thickness of $l_b$ (*31, 71, 92*).[25,67,83] Thus, $R_{int}$ can be described as,

$$R_{int} = \frac{l_b}{D_{O_2,b}} + \frac{H_i R_i}{H_{O_2,b}} \quad \text{(S25)}$$

The first term on the right is the transport resistance in the bulk region, and the second term is the transport resistance in the dense layer region. $H_i$ and $R_i$ are Henry's constant and the diffusion resistance of oxygen in the dense layer, where $R_i$ can be further characterized by the diffusion length $l_i$ and diffusivity $D_i$: $R_i = \frac{l_i}{D_i}$. $H_i$ and $R_i$ are interface-dependent properties and thus the values differ between catalyst-ionomer and catalyst-water interfaces. We'd like to address that $H_i$ and $D_i$ should be distinguished from the bulk properties $H_{O_2,b}$ and $D_{O_2,b}$. Fundamentally, $H_i$ characterizes the oxygen activity at the interface and is governed by the adsorption energy of oxygen at the catalyst-electrolyte interface, and $D_i$ is associated with the energy barrier for oxygen diffusion through the dense electrolyte layer on the catalyst surface, denoted as $E_{ads}$ and $E_a$ respectively. Using the Arrhenius-type equations, $H_i$ and $R_i$ be expressed as,

$$H_i = H_{O_2,b} exp\left(\frac{E_{ads}}{RT}\right) \quad \text{(S26)}$$

$$R_i = \frac{l_i}{D_{O_2,b} exp\left(\frac{-E_a}{RT}\right)} \tag{S27}$$

Thus, the second term on the right side of **equation S25** can be expressed as,

$$\frac{H_i R_i}{H_{O_2,b}} = \frac{l_i}{D_{O_2,b}} exp\left(\frac{E_a + E_{ads}}{RT}\right) \tag{S28}$$

5.3 Key model parameters

At limiting current condition, **equation S22** can be reduced to **equation S29** as $\frac{H_i}{H_{O_2,b} k_r}$ approaches to zero:

$$i_{CL,lim} = \frac{4F\frac{P_{O_2}}{H_{O_2,b}} S_{Pt}}{R_{int}} = \frac{4F\frac{P_{O_2}}{H_{O_2,b}} S_{Pt}}{\frac{l_b}{D_{O_2,b}} + \frac{H_i R_i}{H_{O_2,b}}} = \frac{4F\frac{P_{O_2}}{H_{O_2,b}} S_{Pt}}{\frac{l_b}{D_{O_2,b}} + \frac{l_i}{D_{O_2,b}} exp\left(\frac{E_a + E_{ads}}{RT}\right)} \tag{S29}$$

Thereby, the limiting current is associates with the multiplication of $H_i$ and $R_i$ or the sum of $E_a$ and $E_{ads}$. To describe the catalyst-water interface in the ionomer-free PtNi/PECC CCL, we estimated the value of $(E_a + E_{ads})$ based on the free energy profiles for $O_2$ adsorption on different low-index Pt surfaces in water, determined by MD simulations.[82] The MD simulations show that $(E_a + E_{ads})$ is 0.6 kcal/mol for the Pt (110)-water interface, 1.5 kcal/mol for the Pt (111)-water interface, and 5.1 kcal/mol for the Pt (100)-water interface. Given that Pt (110) and Pt (111) show better $O_2$ adsorption and higher ORR activity, we determined the value of $(E_a + E_{ads})$ as the average of these two facets, yielding a value of 1.05 kcal/mol.

To describe the catalyst-ionomer interface in the conventional ionomer-based CCL, microelectrode measurements from Kudo and colleagues were used to extract the value of $H_i \times R_i$.[84] In their experiments, $H_i \times R_i$ for a Pt-Nafion interface was measured as a function of RH and temperature. The expression below, fitted from experimental data, is used in the model to describe the catalyst-ionomer interface,

$$(H_i \times R_i)_{catalyst-ionomer} = (-530.539 a_0 + 1452.883) \exp\left(\frac{2799[K]}{T}\right) \quad \left[\frac{Pa \cdot m^2 \cdot s}{mol}\right] \tag{S30}$$

where $a_0$ is the water activity in the ionomer solved in the model.

Based on the interfacial parameters described above, parametric studies of electrolyte film thickness were carried out. We assumed $l_i$=0.5 nm as suggested in the MD simulations.[67,82] Values for $l_b$ were varied from 0 to 9.5 nm for the catalyst-ionomer interface and from 0 to 1.5 nm for the catalyst-water interface, summing up to a 0.5-10 nm ionomer film (corresponding to **Supplementary Table 1**)[85-88] and 0.5-2 nm water film based on literature,[66] respectively. Additionally, we calibrated the model with experimental measurements of $R_{NF}$ vs loading for the PtNi/PECC CCL system to ensure that the contribution of Knudsen diffusion to $R_{NF}$ was appropriately captured and maintained constant when comparing between the PtNi/PECC and the conventional ionomer-based CCL.

**Supplementary Table 12** summarizes the key parameters governing local transport resistance used in the model.

**6. Proton transport simulation**

6.1 The estimation of surface sulfonic group density

The density of sulfonic acid on the MWCNT ($\Gamma_{SO_3H}$) is estimated based on **equation S31**:

$$\Gamma_{SO_3H} = \frac{S_{BET} \times M_C}{n_S \times N_A} \quad \text{(S31)}$$

Here, $S_{BET}$ is the BET specific surface area of the CNT (220 $m^2/g$). $M_C$ is the molecular weight of carbon (12 g/mol). $n_S$ is the S atomic ratio to C measured from EDS data (0.38 at%, **Supplementary Table 3**). $N_A$ is Avogadro's number ($6.02 \times 10^{23}\ mol^{-1}$). As a result, $\Gamma_{SO_3H}$ = 1.15 $nm^2$ per sulfonic acid group.

6.2 Simulation Model Configuration Set-Up

In this study, we employed all-atom molecular dynamics (AAMD) simulations to study the surface of PECC, the benzenesulfonate-grafted carbon nanotube systems. For the sake of convenience, however, we used Materials Studio[89] to construct models consisting of benzenesulfonate groups and a 3-layer graphite instead of the ~10 nm multiwall CNT to represent the structure of the PECC support surface, as shown in **Extended Data Fig. 10a**. In our models, we used 16 benzenesulfonate groups arranged in a hexagonal pattern with each sulfonate group separated from the nearest sulfonate groups by 12.3 Å, corresponding to 1.31 $nm^2$ per sulfonic group, which is similar to the density estimation. Once we built the model of the 3-layer graphite surface, we added 16 hydroniums, 224 $H_2O$ molecules, and 413 $O_2$ molecules to the system, corresponding to $\lambda$ = 15 and 200 atm (**Extended Data Fig. 10b,c,** $\lambda$ is the mole ratio of the water and sulfonic group). This λ was based on reported numbers at high-humidity conditions.[90,91] The $O_2$ gas pressure of 200 atm was selected for a good population of $O_2$ molecules in the gas phase for the simulated system.

The bulk Nafion system consists of eight Nafion 117 chains with equivalent weight (EW) = 997 $g/mol_{SO3H}$, similar to Nafion D2020 with EW = 1000 $g/mol_{SO3H}$ used in this project. Each of the chains has a degree of polymerization of 10. The system also contains a total of 1,100 water-related molecules — comprising 1,020 water molecules and 80 hydronium ions — corresponding to a hydration level of $\lambda \sim 14$ (**Extended Data Fig. 10d**). Initial equilibration was performed using an annealing simulation (from 353K to 600K over 600ps for five cycles). This annealing approach was employed to enhance equilibration by temporarily increasing the molecular kinetic energy in the system with more volume, allowing it to overcome local energy minima without introducing structural bias. Details of the annealing protocol are provided in a previous study.[92] Final equilibration involved a 1 ns NVT molecular dynamics (MD) simulation followed by a 600 ns NPT MD simulation, with the final 100 ns used for data analysis.

6.3 Force Fields

During MD simulations, molecular interactions were described using the DREIDING force field[93] with the F3C force field for water[94]. The DREIDING force field has the following form:

$$E_{Total} = E_{vdW} + E_{Q} + E_{Bond} + E_{Angle} + E_{Torsion} + E_{Inversion} \quad \text{(S32)}$$

where $E_{Total}$, $E_{vdW}$, $E_{Q}$, $E_{Bond}$, $E_{Angle}$, $E_{Torsion}$, and $E_{Inversion}$ are total, van der Waals, electrostatic, bond stretching, angle bending, torsion, and inversion energy, respectively. The atomic charge on each molecule was calculated using Mulliken population analysis from Quantum Mechanics (QM).[95] All simulations were performed using Large-scale Atomic/Molecular Massively Parallel Simulator (LAMMPS)[96] MD software. The velocity Verlet algorithm[97] was used to integrate equations of motion, with a time step of 1 fs. The electrostatic interactions were calculated using the particle-particle-particle mesh method (PPPM).[98]

6.4 Constant $O_2$ Pressure Condition.

During the system equilibration, $O_2$ molecules were allowed to diffuse into and out of the PECC surface system dynamically. The $O_2$ molecules were located above the PECC surface system at the beginning. $O_2$ molecules are expected initially to diffuse into the system, reducing the pressure in the gas phase. To maintain the constant pressure condition, we iteratively add $O_2$ molecules into the gas phase every 40 ns until the number of $O_2$ molecules no longer changes significantly. After three iterations, we achieved the constant pressure condition, as summarized in **Supplementary Table 16**. The overall protocol for carrying out an MD simulation is as follows:

(1) First, after constructing the initial configuration, we perform energy minimization, then we heat the system using MD simulation from 10K to 353 K over 600ps, and subsequently, NVT MD simulation at 353 K for 100ns.

(2) In the second step, we check the pressure of the $O_2$ gas phase, add $O_2$ gas molecules until the corresponding gas pressure is attained, and run another 40ns NVT MD simulation.

(3) We iteratively repeat the second step until the pressure does not change significantly. Once we achieve the constant pressure condition at the target pressure value, we perform equilibrium NVT MD simulations for 300 ns.

6.5 Density Profile Analysis.

We employed density profile analysis to determine the distribution of various molecular species in the PECC surface system. To obtain the density profile, we sectionalize the simulated system along the z-axis direction as shown in **Extended Data Fig. 10e** and count the number of atoms in each section to produce the density profiles (**Fig. 5e** and **Extended Data Fig. 10f).**

6.6 Simulation of effective vehicular proton diffusivity of IonoSkin or Nafion (proton conductors).

We used “proton conductors” as a general name to simplify the description. It refers to the IonoSkin in the context of the PECC CCL and the Nafion in the context of the conventional CCL.

Here, we define the *effective* vehicular proton diffusivity as the proton diffusivity that accounts for the influence of the water–hydronium network (WHN) boundaries, which confine proton transport within the connected hydrophilic domains. This definition better represents long-range vehicular diffusion in pure proton conductors. The term "effective" is used to distinguish it from the unrestricted, or *bulk*, vehicular proton diffusivity, which describes proton transport in an unconfined, infinite aqueous phase.

To analyze the effective vehicular proton diffusion coefficient, we calculated the mean-square displacement (MSD) of hydronium during the equilibrium state of our simulation using the equation:

$$MSD = \frac{1}{N}\sum_{i=1}^{N}|\boldsymbol{r}_i(t) - \boldsymbol{r}_i(0)|^2 \tag{S33}$$

where $\boldsymbol{r}_i(t)$ and $\boldsymbol{r}_i(0)$ denote the position of the particle $i$ at time $t$ and the beginning, which is restricted within the WHN. $N$ denotes the number of hydronium molecules. From the MSD, the effective vehicular diffusion coefficient ($D_{eff-Vehicular}$) was evaluated:

$$D_{eff-Vehicular} = \frac{1}{2d}\lim_{t\to\infty}\frac{1}{t}MSD = \frac{1}{2dN}\lim_{t\to\infty}\frac{1}{t}\sum_{i=1}^{N}|\boldsymbol{r}_i(t) - \boldsymbol{r}_i(0)|^2 \tag{S34}$$

$d$ is the number of dimensions. For the IonoSkin, the $d = 2$. For Nafion, the $d = 3$. We note that the $D_{Vehicular}$ was derived from the last 100 ns long segments from the 300 ns NVT equilibrium MD simulations for statistical analysis, which is a long-time (Fickian) regime. Thus, the vehicular diffusivity already reflects the effect of the WHN boundary.

When only one spatial dimension component of the $D_{eff-Vehicular}$ is considered (e.g., $D_x$), the position vector $r_i(t)$ collapses to its scalar component along that direction, (e.g., $x_i(t)$). Correspondingly, the denominator in **equation S34** reduces to 2Nt.

6.7 Calculation of the water-hydronium network (WHN) tortuosity factor ($\tau_{WHN}$) of proton conductors

We note that the tortuosity factor can be confused with tortuosity since both are commonly represented by τ. Since the tortuosity factor is the square of the tortuosity, this confusion may lead to mistakes, as has been noted in literature.[99] In this manuscript, we only discuss the tortuosity factor. τ is used to represent the tortuosity factor, following the terminology in other ionomer morphology studies.

The water-hydronium network (WHN) tortuosity factor ($\tau_{WHN}$) of proton conductors can be quantified from the ratio of vehicular diffusivities obtained in bulk acidic solution, which represents an unrestricted vehicular transport, relative to the effective vehicular diffusivity in proton conductors:

$$\tau_{WHN} = \frac{D_{Vehicular-Bulk\ acidic\ solution}}{D_{eff-Vehicular}} \tag{S35}$$

where $D_{Vehicular-Bulk\ acidic\ solution}$ is the vehicular proton diffusivity in the bulk acidic solution system, $D_{eff-Vehicular}$ is the effective vehicular proton diffusivity in proton conductors obtained from **equation S34**.

$D_{Vehicular-Bulk\ acidic\ water}$ was obtained from a separate simulation of a 3.7M HCl aqueous solution under identical temperature conditions (353 K). The simulated bulk acidic water system is composed of 840 water molecules, 60 hydronium ions, and 60 Cl counter ions, corresponding to λ = 15 and a hydronium concentration of 3.7 M. A similar annealing and heating protocol to that applied to bulk Nafion was applied to equilibrate the bulk water system. Afterwards, a 10 ns NPT simulation was conducted as the final equilibration stage, and the last 5 ns was used for diffusion analysis. Vehicular Proton diffusion of this bulk acidic water system was found to be $D_{Vehicular-Bulk\ acidic\ solution} = 1.54 \times 10^{-5}\ cm^2/s$.

6.8 Simulation of effective hopping proton diffusivity of IonoSkin or Nafion (proton conductors).

To analyze the effective hopping proton diffusivity, we first employed the method developed by Deng et al. [100,101]to calculate the bulk hopping diffusivity, which excludes the WHN boundary effect. The bulk diffusivity for the hopping mechanism is computed using **equation S36**:

$$D_{Bulk-Hopping} = \frac{1}{2dNt}\int_0^{t\to\infty} \sum_i^N \sum_i^M k_{ij} r_{ij}^2 P_{ij}\, dt \tag{S36}$$

where $d$ is the number of dimensions ($d = 2$ for the IonoSkin, $d = 3$ for Nafion), $N$ is the number of protons, $r_{ij}$ is the distance between all donors and acceptors measured from the simulations in the equilibrium state. $P_{ij}$ is the probability that a proton will jump from hydronium i to water molecule j, defined as $P_{ij} = \frac{k_{ij}}{\sum_j^M k_{ij}}$ where $k_{ij}$ is defined as:

$$k_{ij}(r) = \kappa(T,r)\frac{k_b T}{h}\exp\left(-\frac{E_{ij}(r)-\frac{1}{2}h\omega(r)}{RT}\right) \tag{S37}$$

where $\kappa(T,r)$ is the tunneling factor, $\omega(r)$ is the frequency for the zero-point energy correction,[102,103] $k_b$ is Boltzmann constant, $T$ is the temperature, $h$ is Planck's constant, $E_{ij}(r)$ is the energy barrier for a proton to hop from donor to acceptor as a function of the distance $(r)$ between them. The hopping energy barrier was assessed using the description by Jang et al.[104]

The $D_{Bulk-Hopping}$ reflects only the exchange between a hydronium ion and its immediate water neighbors without considering the confinement of WHN boundaries. Thus, the effective hopping diffusivity can be obtained by **equation S38,** which considers the tortuosity of the WHN:[69,105]

$$D_{eff-Hopping} = \frac{1}{\tau_{WHN}} D_{Bulk-Hopping} \quad \text{(S38)}$$

Here, the $\tau_{WHN}$ is obtained from **equation S35**.

6.9 Calculation of the effective total proton diffusivity

The effective total proton diffusivity $(D)$ is the sum-up of the corresponding vehicular and hopping diffusivity:

$$D = D_{eff-Vehicular} + D_{eff-Hopping} \quad \text{(S39)}$$

where $D_{eff-Vehicular}$ and $D_{eff-Hopping}$ are the vehicular and hopping diffusivities obtained from **equations S34** and **S38**.

6.10 Calculation of hydronium concentration in proton conductors

The hydronium concentration in the IonoSkin of the PECC surface system $(c_{IonoSkin})$ was calculated based on the depth profile between 4 Å to 9 Å from the outermost graphene layer.

$$c_{IonoSkin} = \frac{N_{IonoSkin}}{V_{IonoSkin} \times N_A} \quad \text{(S40)}$$

Here, $N_{IonoSkin}$ is the total number of hydroniums in the proton layer based on the depth profile between 4 Å to 9 Å ($N_{IonoSkin}$ = 16). $V_{IonoSkin}$ is the volume of the proton layer calculated by the box length (49.2 Å), box width (42.6 Å), and the thickness of the water layer (5 Å). As a result, the $c_{IonoSkin} = 2.53 \times 10^{-3}\ mol/cm^3$.

The hydronium concentration in the Nafion of conventional CCL $(c_{Nafion})$ was calculated based on

$$c_{Nafion} = \frac{N_{Nafion}}{V_{Nafion} \times N_A} \quad \text{(S41)}$$

Here, $N_{Nafion}$ is the total number of hydroniums in the bulk Nafion ($N_{Nafion}$ = 80). $V_{Nafion}$ is the volume of the bulk Nafion calculated by the box length (46.7 Å), box width (46.7 Å), and box height (46.7 Å). As a result, the $c_{Nafion} = 1.31 \times 10^{-3}\ mol/cm^3$.

6.11 Calculation of proton conductivity of proton conductors

We calculated the proton conductivity based on the hydronium diffusion using a modified Nernst-Einstein equation:[69]

$$\sigma = \frac{z^2F^2}{RT} \times D \times c_{Proton\ conductor} \tag{S42}$$

In **equation S42**, σ is the proton conductivity of the proton conductor. $D$ is the effective total proton diffusivity obtained from **equation S39**. $c_{Proton}$ is the concentration of hydronium ions in the proton conductors that is calculated by **equations S40** and **S41**. z is the charge of the conducting ion (1.0 for hydronium). F is the Faraday constant, R is the gas constant, and T is the simulation temperature (353 K). As a result, the $\sigma_{IonoSkin} = 28.8\ S/m$, the $\sigma_{Nafion} = 8.26\ S/m$

6.12 Estimation of the proton conductor volume fraction in the CCL

In the PECC CCL, we estimated the IonoSkin volume fraction ($\varepsilon_{IonoSkin}$) based on **equation S43**:

$$\varepsilon_{IonoSkin} = \frac{V_{IonoSkin}}{V_{CCL}} = \frac{Unit\ area \times m_c \times S_{BET} \times t_{IonoSkin}}{Unit\ area \times t_{CCL}} \tag{S43}$$

Here, the $V_{IonoSkin}$ is the volume of the IonoSkin in the CCL. $V_{CCL}$ is the volume of the whole CCL. $m_c$ is the area-normalized mass loading of carbon ($m_c$ = 0.36 mg/cm$^2$ for CCLs with $m_{Pt}$ = 0.09 mg/cm$^2$). $S_{BET}$ is the BET specific surface area of the CNT (220 m$^2$/g). $t_{IonoSkin}$ is the thickness of the proton layer (5 Å). $t_{CCL}$ is the thickness of the proton layer (5.4 µm). As a result, the $\varepsilon_{IonoSkin}$ = 7.33 vol%.

In the conventional CCL (c-PtNi/C+I), we estimated the Nafion volume fraction ($\varepsilon_{Nafion}$) based on **equation S44**:

$$\varepsilon_{Nafion} = \frac{V_{Nafion}}{V_{CCL}} = \frac{Unit\ area \times m_c \times IC\ ratio}{Unit\ area \times t_{CCL} \times \rho_{Ionomer}} \tag{S44}$$

Similarly, the $V_{Proton\ layer}$ is the volume of the proton layer in the CCL. $V_{CCL}$ is the volume of the whole CCL. $m_c$ is the area-normalized mass loading of carbon ($m_c$ = 0.36 mg/cm$^2$ for CCLs with $m_{Pt}$ = 0.09

mg/cm$^2$). $IC\ ratio$ is the ionomer/carbon weight ratio in the conventional CCL ($IC\ ratio$ = 0.8). $t_{CCL}$ is the thickness of the proton layer (6.2 μm). $\rho_{Ionomer}$ is the ionomer density at λ = 14 (1.7 g/cm$^3$ based on literature.[106] As a result, the $\varepsilon_{Nafion\ in\ conventional\ CCL}$ = 27.3 vol%.

6.13 Calculation of the tortuosity factor of proton conductor distribution in the CCL ($\tau_{CCL}$)

The $\tau_{CCL}$ was calculated based on **equation S45** obtained from the literature.[29]

$$\tau_{CCL} = \frac{\varepsilon_{Proton\ conductor} \times \sigma_{Proton\ conductor}}{\sigma_{CCL}} \quad \text{(S45)}$$

Here, $\varepsilon_{Proton\ conductor}$ is the proton conductor volume fraction in the CCL. $\varepsilon_{IonoSkin} = 7.33\ vol\%$ and $\varepsilon_{Nafion} = 27.3\ vol\%$. $\sigma_{Proton\ conductor}$ is the proton conductivity of the proton conductor. $\sigma_{IonoSkin} = 28.8\ S/m$ and $\sigma_{Nafion} = 8.26\ S/m$. $\sigma_{CCL}$ is the experimental proton conductivity of the CCL. $\sigma_{TIF\ CCL} = 1.19\ S/m$ and $\sigma_{Conventional\ CCL} = 1.14\ S/m$. As a result, $\tau_{CCL,IonoSkin} = 1.77$, $\tau_{CCL,Nafion} = 1.98$.

## Data availability:

The data that support the major findings of this study are available in the main text or the Supplementary Information. Data on the MD simulation of proton transport in the main text that support the findings of this study will be available via a publicly accessible depository. Further data are available from the corresponding authors upon reasonable request.

## Reference

72 Garrick, T. R., Moylan, T. E., Yarlagadda, V. & Kongkanand, A. Characterizing Electrolyte and Platinum Interface in PEM Fuel Cells Using CO Displacement. *J. Electrochem. Soc.* **164**, F60-F64, doi:10.1149/2.0551702jes (2016).

73 Qi, Y. *et al.* Understanding Platinum Ionomer Interface Properties of Polymer Electrolyte Fuel Cells. *J. Electrochem. Soc.* **169**, 064512, doi:10.1149/1945-7111/ac774f (2022).

74 Jackson, C. *et al.* Support induced charge transfer effects on electrochemical characteristics of Pt nanoparticle electrocatalysts. *J. Electroanal. Chem.* **819**, 163-170, doi:10.1016/j.jelechem.2017.10.010 (2018).

75 Xia, Z. & Chan, S. Analysis of carbon-filled gas diffusion layer for H2/air polymer electrolyte fuel cells with an improved empirical voltage–current model. *Int. J. Hydrogen Energy* **32**, 878-885, doi:10.1016/j.ijhydene.2006.12.013 (2007).

76 Morris, D. R., Liu, S. P., Villegas Gonzalez, D. & Gostick, J. T. Effect of water sorption on the electronic conductivity of porous polymer electrolyte membrane fuel cell catalyst layers. *ACS Appl. Mater. Interfaces* **6**, 18609-18618, doi:10.1021/am503509j (2014).

77 Tongsh, C. *et al.* Fuel cell stack redesign and component integration radically increase power density. *Joule* **8**, 175-192, doi:10.1016/j.joule.2023.12.003 (2024).

78 Yoshizumi, T., Kubo, H. & Okumura, M. Development of High-Performance FC Stack for the New MIRAI. Report No. 2021-01-0740, (SAE International, 2021). Available at https://saemobilus.sae.org/papers/development-high-performance-fc-stack-new-mirai-2021-01-0740.

79 Wang, G., Osmieri, L., Star, A. G., Pfeilsticker, J. & Neyerlin, K. C. Elucidating the Role of Ionomer in the Performance of Platinum Group Metal-free Catalyst Layer via in situ Electrochemical Diagnostics. *J. Electrochem. Soc.* **167**, 044519, doi:10.1149/1945-7111/ab7aa1 (2020).

80 Sui, P.-C., Zhu, X. & Djilali, N. Modeling of PEM Fuel Cell Catalyst Layers: Status and Outlook. *Electrochem. Energy Rev.* **2**, 428-466, doi:10.1007/s41918-019-00043-5 (2019).

81 Pant, L. M. *et al.* Along-the-channel modeling and analysis of PEFCs at low stoichiometry: Development of a 1+2D model. *Electrochim. Acta* **326**, 134963, doi:10.1016/j.electacta.2019.134963 (2019).

82 Wang, S., Zhu, E., Huang, Y. & Heinz, H. Direct correlation of oxygen adsorption on platinum-electrolyte interfaces with the activity in the oxygen reduction reaction. *Sci. Adv.* **7**, eabb1435, doi:doi:10.1126/sciadv.abb1435 (2021).

83 Mann, R. F., Amphlett, J. C., Peppley, B. A. & Thurgood, C. P. Henry's Law and the solubilities of reactant gases in the modelling of PEM fuel cells. *J. Power Sources* **161**, 768-774, doi:10.1016/j.jpowsour.2006.05.054 (2006).

84 Kudo, K., Jinnouchi, R. & Morimoto, Y. Humidity and Temperature Dependences of Oxygen Transport Resistance of Nafion Thin Film on Platinum Electrode. *Electrochim. Acta* **209**, 682-690, doi:10.1016/j.electacta.2016.04.023 (2016).

85 Lopez-Haro, M. *et al.* Three-dimensional analysis of Nafion layers in fuel cell electrodes. *Nat. Commun.* **5**, 5229, doi:10.1038/ncomms6229 (2014).

86 Inoue, G. *et al.* Simulation of carbon black aggregate and evaluation of ionomer structure on carbon in catalyst layer of polymer electrolyte fuel cell. *J. Power Sources* **439**, 227060, doi:10.1016/j.jpowsour.2019.227060 (2019).

87 So, M., Park, K., Tsuge, Y. & Inoue, G. A Particle Based Ionomer Attachment Model for a Fuel Cell Catalyst Layer. *J. Electrochem. Soc.* **167**, 013544, doi:10.1149/1945-7111/ab68d4 (2020).

88 Park, K. *et al.* Evaluation of ionomer distribution on porous carbon aggregates in catalyst layers of polymer electrolyte fuel cells. *J. Power Sources Adv.* **15**, doi:10.1016/j.powera.2022.100096 (2022).

89 Materials Studio v. 5.5 (Accelrys Software Inc., San Diego, CA, 2010).

90 Zawodzinski, T. A. *et al.* Water Uptake by and Transport Through Nafion® 117 Membranes. *J. Electrochem. Soc.* **140**, 1041, doi:10.1149/1.2056194 (1993).

91 Choi, P. & Datta, R. Sorption in Proton-Exchange Membranes: An Explanation of Schroeder's Paradox. *J. Electrochem. Soc.* **150**, E601, doi:10.1149/1.1623495 (2003).

92 Bazaid, M., Huang, Y., Goddard, W. A. & Jang, S. S. Proton transport through interfaces in nanophase-separation of hydrated aquivion membrane: Molecular dynamics simulation approach. *Colloids Surf. A: Physicochem. Eng. Asp.* **676**, 132187, doi:10.1016/j.colsurfa.2023.132187 (2023).

93 Mayo, S. L., Olafson, B. D. & Goddard, W. A. DREIDING - A generic Force-Field for molecular simulations. *J. Phys. Chem.* **94**, 8897-8909, doi:10.1021/j100389a010 (1990).

94 Levitt, M., Hirshberg, M., Sharon, R., Laidig, K. E. & Daggett, V. Calibration and testing of a water model for simulation of the molecular dynamics of proteins and nucleic acids in solution. *J. Phys. Chem. B* **101**, 5051-5061, doi:10.1021/jp964020s (1997).

95 Mulliken, R. S. Electronic Population Analysis on LCAO–MO Molecular Wave Functions. I. *J. Chem. Phys.* **23**, 1833-1840, doi:10.1063/1.1740588 (1955).

96 Thompson, A. P. *et al.* LAMMPS-a flexible simulation tool for particle-based materials modeling at the atomic, meso, and continuum scales. *Comput. Phys. Commun.* **271**, 34, doi:10.1016/j.cpc.2021.108171 (2022).

97 Verlet, L. Computer "Experiments" on Classical Fluids. I. Thermodynamical Properties of Lennard-Jones Molecules. *Phys. Rev.* **159**, 98-103, doi:10.1103/PhysRev.159.98 (1967).

98 Hockney, R. W. & Eastwood, J. W. *Computer Simulation Using Particles*. 1st edn, 267–304 (CRC Press, 1988).

99 Johansson, E. O., Yamada, T., Sundén, B. & Yuan, J. Dissipative particle dynamics approach for nano-scale membrane structure reconstruction and water diffusion coefficient estimation. *Int. J. Hydrogen Energy* **40**, 1800-1808, doi:10.1016/j.ijhydene.2014.11.030 (2015).

100 Jang, S. S., Lin, S.-T., Çağın, T., Molinero, V. & Goddard, W. A. Nanophase Segregation and Water Dynamics in the Dendrion Diblock Copolymer Formed from the Fréchet Polyaryl Ethereal Dendrimer and Linear PTFE. *The Journal of Physical Chemistry B* **109**, 10154-10167, doi:10.1021/jp050125w (2005).

101 Deng, W.-Q., Molinero, V. & Goddard, W. A. Fluorinated Imidazoles as Proton Carriers for Water-Free Fuel Cell Membranes. *J. Am. Chem. Soc.* **126**, 15644-15645, doi:10.1021/ja046999y (2004).

102 Lill, M. A. & Helms, V. Compact parameter set for fast estimation of proton transfer rates. *J. Chem. Phys.* **114**, 1125-1132, doi:10.1063/1.1332993 (2001).

103 Sone, Y., Ekdunge, P. & Simonsson, D. Proton Conductivity of Nafion 117 as Measured by a Four-Electrode AC Impedance Method. *J. Electrochem. Soc.* **143**, 1254, doi:10.1149/1.1836625 (1996).

104 Brunello, G., Lee, S. G., Jang, S. S. & Qi, Y. A molecular dynamics simulation study of hydrated sulfonated poly(ether ether ketone) for application to polymer electrolyte membrane fuel cells: Effect of water content. *Journal of Renewable and Sustainable Energy* **1**, doi:10.1063/1.3138922 (2009).

105 Thieu, L. M., Zhu, L., Korovich, A. G., Hickner, M. A. & Madsen, L. A. Multiscale Tortuous Diffusion in Anion and Cation Exchange Membranes. *Macromolecules* **52**, 24-35, doi:10.1021/acs.macromol.8b02206 (2018).

106 Kusoglu, A. & Weber, A. Z. New Insights into Perfluorinated Sulfonic-Acid Ionomers. *Chem. Rev.* **117**, 987-1104, doi:10.1021/acs.chemrev.6b00159 (2017).

## Acknowledgement:

We acknowledge the use of facilities and instrumentation at the UC Irvine Materials Research Institute (IMRI). We also thank the Electron Imaging Center of Nanomachines (EICN) at California NanoSystems Institute (CNSI) for TEM and SEM support. We acknowledge the training and help from I. Martini for the XPS data collection at the UCLA Molecular Instrumentation Center (MIC).

## Funding:

Y.H. and A.Z. acknowledge partial support from the National Science Foundation under award 2404462 and the Defense University Research Instrumentation Program (DURIP) grant N00014-18-1-2271. W.A.G acknowledges support by National Science Foundation (NSF) grant CBET 2311117. S.W. and A.W. acknowledge support by the Million Mile Fuel Cell Truck (M2FCT) Consortium (https://millionmilefuelcelltruck.org) (technology manager: Greg Kleen), which is supported by the U.S. Department of Energy, Office of Energy Efficiency and Renewable Energy, Hydrogen and Fuel Cell Technologies Office, under contract number DE-AC02–05CH1123.

**Author Contributions**

Research design and conceptualization: Y.H. and A.Z.; Experimental design and execution: A.Z., R.W.; Synthesis of electrocatalysts: A.Z., R.W.; MEA fabrication, and electrochemical testing: A.Z., R.W.; Characterizations: A.Z., R.W., Y.-H.J.T., B.P., Z.L., B.Z., and T.-J.H.; MD simulation of proton transport: M.O.B.; Multi-physics model of oxygen transport: S.W.; EIS analysis and fitting of proton transport: A.Z. and A.S.. Supervision: Y.H. for all experiments; W.A.G. and S.S.J. for MD simulation of proton transport; A.W. for Multiphysics model of oxygen transport. Writing: original draft: A.Z., M.O.B., S.W., Y.H., Y.W.; review & editing: Y.H., W.A.G., S.S.J., X.D., and A.W. All authors approved the final version of the manuscript.

**Competing interests:**

The UCLA team has submitted a patent application by the UCLA inventors: Y.H. and A.Z.; the remaining authors declare no competing interests.